\documentclass{svjour3}                     

\smartqed 
\usepackage[utf8]{inputenc}
\usepackage{amsmath}
\usepackage{amsfonts}
\usepackage{amssymb}
\usepackage{listings}
\usepackage{mathrsfs}
\usepackage{times}
\usepackage{float}
\usepackage{color}
\usepackage{setspace}
\usepackage{color,soul}

\usepackage{amsmath,amsfonts,amsthm,eucal, bm}
\usepackage{enumerate}
\usepackage[letterpaper,margin=3cm]{geometry}
\usepackage{float}
\usepackage[title]{appendix}
\usepackage{color}
\usepackage{comment}
\usepackage{multirow}
\usepackage{caption}
\usepackage{enumitem}
\usepackage{longtable}
\usepackage{fancybox}

\usepackage[
pdfstartview=XYZ,
bookmarks=true,
colorlinks=true,
linkcolor=blue,
urlcolor=blue,
citecolor=blue,
pdftex,
bookmarks=true,
linktocpage=true,   
hyperindex=true
]{hyperref}

\usepackage{lipsum}
\usepackage{lineno}
\usepackage[FIGBOTCAP,TABTOPCAP,bf,tight]{subfigure}

\usepackage[numbers, sort&compress]{natbib}

\usepackage{graphicx}
\usepackage{pstool}
\usepackage{psfrag}
\usepackage{epstopdf}
\usepackage{booktabs} 
\usepackage{lineno}
\usepackage{import}
\usepackage[export]{adjustbox}

\graphicspath{{figures/}}

\newcommand\gbf[1]{\hbox{\boldmath${#1}$\unboldmath}}

\DeclareMathOperator{\tr}{tr}

\renewcommand{\div} {\mbox{div}}		

\newcommand  {\grad}{\mbox{grad}}

\newcommand{\rmd}{{\rm d}}
\newcommand{\Ds}{\displaystyle}	
\newcommand{\aR}[0]{{\alpha\!R}}
\newcommand{\RhoaR}	{\rho^\aR}
\newcommand{\FR}[0]{{F\!R}}

\newcommand{\p}[0]{\partial\hspace*{1pt}}

\newcommand{\Teps}{\hbox{{\large\boldmath$\varepsilon$\unboldmath}}}

\newcommand{\Vx}{{\mathbf{x}}}
\newcommand{\VX}{{\mathbf{X}}}
\newcommand{\Vv}{{\mathbf{v}}}
\newcommand{\Vu}{{\mathbf{u}}}

\newcommand{\TT}{{\mathbf{T}}}
\newcommand{\TK}{{\mathbf{K}}}
\newcommand{\TI}{{\mathbf{I}}}
\newcommand{\TF}{{\mathbf{F}}}

\newcommand{\Vg}{{\mathbf{g}}}

\newcommand{\Vp}{{\mathbf{p}}}
\newcommand{\Ve}{{\mathbf{e}}}

\newcommand{\Bhat}[1]{{\bf\hat{\mbox{$#1$}}}{}}

\newcommand\earg[0]{\hbox{$(\,\hbox{\newbullet}\,)$}}
\newcommand{\Vnull}{\mbox{\bf 0}}
\def\newbullet{%
  \unitlength 1ex%
  \begin{picture}(1,1)
  \put(0.5,0.6){\circle*{0.4}}
  \end{picture}}

 \newcommand{\Vw}{{\mathbf{w}}}

\newcommand{\Half}[0]{\mbox{$\frac{1}{2}$}}

\usepackage{algorithm}
\usepackage[noend]{algpseudocode}

\newcommand{\NAME}[1]{{\slshape{#1}\/}}

\newcommand{\Lame}     {\NAME{Lam\'{e}}}

\newcommand{\Vxi}  {\gbf{\xi}}
\newcommand{\Vchi}  {\raisebox{0.3ex}{\footnotesize$\gbf{\chi}$}}

\newcommand{\Vxs}{\mbox{$\stackrel{\prime}{\Vx}$}}

\newcommand{\Sum}{\sum\limits}			
\newcommand{\Suma}{\Sum_{\raisebox{1ex}{$\scriptstyle\alpha$}}}

\journalname{My-journal}
\begin{document}

\title{A multiscale CNN-based intrinsic permeability prediction in deformable porous media}

\author{Yousef Heider$^{1,\ast}$
        \and
        Fadi Aldakheel$^{1}$
        \and
        Wolfgang Ehlers$^{2}$
}

\institute{ Y. Heider\at
              Institute of General Mechanics, RWTH Aachen University, Germany \\
              \email{heider@iam.rwth-aachen.de}}

\institute{$^1$Institute of Mechanics and Computational Mechanics (IBNM), Leibniz University Hannover.\\
Appelstr. 9A, 30167 Hannover, Germany\\[1mm]
$^2$ Institute of Applied Mechanics, University of Stuttgart.\\
Pfaffenwaldring 7, 70569 Stuttgart, Germany\\[1mm]
$^{\ast}$ Corresponding author: Yousef Heider.\\ 
\email{yousef.heider@ibnm.uni-hannover.de}
}
                 
\date{Received: date / Accepted: date}

\maketitle

\begin{abstract}   

This work introduces a novel application for predicting the macroscopic intrinsic permeability tensor in deformable porous media, using a limited set of micro-CT images of real microgeometries. The primary goal is to develop an efficient, machine-learning (ML)-based method that overcomes the limitations of traditional permeability estimation techniques, which often rely on time-consuming experiments or computationally expensive fluid dynamics simulations. The novelty of this work lies in leveraging Convolutional Neural Networks (CNN) to predict pore-fluid flow behavior under deformation and anisotropic flow conditions. 
Particularly, the described approach employs binarized CT images of porous micro-structure as inputs to predict the symmetric second-order permeability tensor, a critical parameter in continuum porous media flow modeling. The methodology comprises four key steps: (1) constructing a dataset of CT images from Bentheim sandstone at different volumetric strain levels; (2) performing pore-scale simulations of single-phase flow using the lattice Boltzmann method (LBM) to generate permeability data; (3) training the CNN model with the processed CT images as inputs and permeability tensors as outputs; and (4) exploring techniques to improve model generalization, including data augmentation and alternative CNN architectures.
Examples are provided to demonstrate the CNN’s capability to accurately predict the permeability tensor, a crucial parameter in various disciplines such as geotechnical engineering, hydrology, and material science.
An exemplary source code is made available for interested readers [\href{https://doi.org/10.25835/xrii0m6f}{https://doi.org/10.25835/xrii0m6f}].

 
\end{abstract}

\begin{center}
  \fbox{ 
    \begin{minipage}{0.9\textwidth}
    \textbf{Abbreviations} 
    \begin{longtable}{ll ll}
    ANN & Artificial Neural Networks & CNN & Convolutional Neural Networks \\
    CT & Computed Tomography & RVE & Representative Volume Elements \\
    LBM & Lattice Boltzmann Method & TPM & Theory of Porous Media \\
    SSA & Specific Surface Area & ML & Machine Learning \\
    BGK & Bhatnagar-Gross-Krook & DNN & Deep Neural Networks \\
    GAN & Generative Adversarial Networks & PNM & Pore Network Model \\
    RNN & Recurrent Neural Networks & FFNN &Feed-Forward  Neural Networks \\
    \end{longtable}
    \end{minipage}
  }
\end{center}

\section{Introduction}
\label{intro}

The study of single and multiphase flow in deformable and possibly fractured porous materials is a topic of interest in various fields, such as petroleum engineering, hydrology, soil science, and biomechanics. 
Related applications and references can be found in, e.g., \citet{Ehlers2022_DarcyForchheimerBrinkman, EhlersEtAl2004_UnsaturatedPM, MosthafEtAl2011_FreeFlow_PM, Maea10, Markert2013, EhlersLuo17, EhlersWagner2019, HeiderSun2020PFMunsaturated, Heider2021_review_PFHyd, PetersEtAl2023_PAMM, MieheMauthe16_part3_porous, GAWIN2020322_PM_Freezing, wang2019updated, choo2015stabilized, jenny2005adaptive}.
The understanding and accurate mathematical description of fluid flow in these media is crucial for optimizing extraction processes, predicting contaminant transport, improving soil irrigation techniques, and predicting the stability of geotechnical structures, among others. 
However, the inherent heterogeneity and anisotropy of porous media pose a major challenge to the accurate prediction of their macroscopic properties, such as permeability. This underlines the need to incorporate the microscopic information through an accurate but computationally efficient multiscale approach.

In recent years, machine learning (ML), particularly Artificial Neural Networks (ANN) has emerged as a promising tool to tackle several problems in multi-scale material modeling. 
Using, e.g., Deep Neural Networks (DNN), constitutive models based on lower-scale simulations can be found in different research works, such as within crystal plasticity \cite{heider2020so, HeiderSun2023_PAMM_PlasObj}, elasto-plasticity for multiphase, composite materials \cite{Fuchs2021_DNN2_RL}, fracture mechanics \cite{tragoudas2024enhanced,aldakheel2021feed}, hyperelasticity with enforced constitutive restrictions, as the symmetry of the stress tensor, objectivity, material symmetry, polyconvexity, and thermodynamic consistency \cite{LindenKaestner2023_HyperElasticity}, 
and other applications in \cite{Wessels2022}. A recent review on ML applications within solid mechanics is presented by \citet{HanxunEtAl2023_ML-Review}.
Within multi-scale modeling of porous materials, ANN, including Convolutional Neural Networks (CNN) and regression models, have demonstrated their ability to learn complex patterns and relationships. This made them suitable for predicting macroscopic properties and material models, such as the anisotropic permeability, the retention curves of unsaturated porous materials, and the inelastic stress-strain relationships, see, e.g. \cite{wang2018multiscale, HeiderHSSuh2020_ML_offline,CaiEtAl_2023_GeomLear, ChaabanEtAl2023_ML_LBM, WU2018CNN_Permeability_Synthatic,HongLiu2020_CNN_Permeability} for an overview. 
Specifically, CNN approach holds significant capabilities in linking information across scales and in model parameter prediction within computational mechanics as highlighted in these two review papers \citet{BisharaEtAl2023_Review_ML_Multiscale, Herrmann_Kollmannsberger_2024} as examples.
The application of ML in this context has the potential to significantly reduce the computational costs associated with traditional multiscale simulation methods while ensuring a higher degree of accuracy in comparison with the phenomenological models.
However, the good performance of these surrogate models requires a sufficient amount of data, e.g., small-scale CT images or numerical data. 
If the available image database is small, data enhancement techniques can help to overcome this challenge. For example, \citet{NguyenEtAl2022_GAN-RL_Sun} have used generative adversarial networks (GAN) together with actor-critic (AC) reinforcement learning to synthesize realistic and controlled 3D microstructures, which are also used in this work.

For the macroscopic simulation of multiphase, heterogeneous, and deformable porous materials, the Theory of Porous Media (TPM) can be employed, which represents a reliable and robust framework, as has been demonstrated in many works, e.g. \cite{Ehle02, EhlersEtAl2004_UnsaturatedPM, markert2007constitutive, HEIDER2014_Liquifaction, EhlersWagner2019, SweidanEtAl2019, SweidanEtAl020_Unified_Kinematics_CMAME,chaaban2020upscaling, markert2011advances, ArchutEtAl2022_Fiber_ITA_IAM_IGMR, DeMarchi_EtAl_2024_PM}. 
In this context, important parameters, such as the deformation-dependent intrinsic permeability, the relative permeabilities, and the degree of saturation with relation to the capillary pressure determine the flow behavior in the continuum multiphase porous media models. 
Alternative to the phenomenological models and their assumptions, these parameters and the associated constitutive formulations can be estimated based on lower-scale flow simulations. For instance, the lattice Boltzmann method (LBM) can be applied to simulate the flow on the microscale through representative volume elements (RVEs) of the analyzed material. For model details and references, see, e.g., \cite{wang2019updated, chaaban2020upscaling, chaaban2022_TwoPhaseLBMtpm,PhuEtAl2023_PAMM}.

%

\begin{figure}[!ht]
\begin{center}
\includegraphics[width=14.0cm]{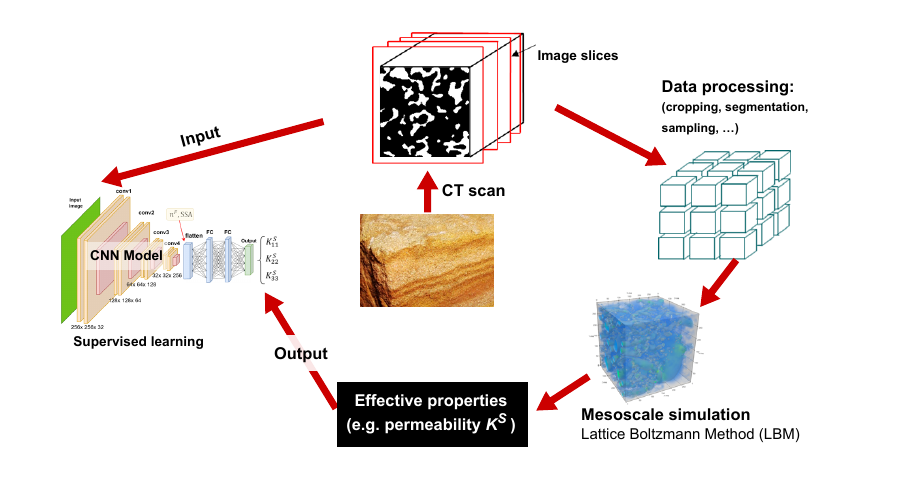}
\end{center}
\caption{Overview of the major steps of the 2D CNN model to predict porous media effective properties (mainly the permeability) based on CT images. This includes the input, the output, dataset preparation, and model training.}
\label{Fig:Overview}
\end{figure}

Within ML-based multiscale modeling of flow through porous materials, regression models using simple Feed-Forward  Neural Networks (FFNN) are successfully utilized by \citet{HeiderHSSuh2020_ML_offline} and \citet{ChaabanEtAl2023_ML_LBM} for capturing the time-independent  permeability tensor. On the other hand, path-dependent responses, such as the retention model, could be captured using the Recurrent Neural Networks (RNN) or the one-dimensional (1D) Convolutional Neural Networks (1D-CNN) as thoroughly discussed in \cite{HeiderHSSuh2020_ML_offline, ChaabanEtAl2023_ML_LBM, Heider2021_Habil}.
While the aforementioned regression models use only the LBM datasets, in this work we discuss the use of two-dimensional (2D) convolutional neural networks (CNN) to predict the macroscopic anisotropic permeability tensor at different deformation states. 
The supervised learning process in this context uses binarized CT images of real microgeometry as inputs, while the output is the 2$^\text{nd}$-order symmetric intrinsic permeability tensor. 
Figure~\ref{Fig:Overview} illustrates the major steps in this model, which include building the dataset based on  CT images from Bentheim sandstone and training the 2D CNN model with the processed CT images as inputs and permeability tensors as outputs.
This CNN-based treatment has shown its effectiveness in various studies, such as within multiscale magnetostatics \cite{Aldakheel_2023_CNN_Magnetostatics} and multiscale modeling of heterogeneous materials \cite{Aldakheel_2023_CNN_Hetero}.

After this introduction, the rest of the paper is organized as follows: Section~\ref{sec:theoryContinum} presents the fundamentals of macroscopic modeling of saturated porous media based on the TPM. This section covers the concepts of homogenization, kinematics, local balance relations, and essential constitutive relations for biphasic porous materials. 
Additionally, it reviews modeling the flow through porous media, including Darcy's and Darcy-Brinkman’s models \cite{Ehlers2022_DarcyForchheimerBrinkman}.
This is followed in Section~\ref{sec:Database} by a discussion of the data generation steps.
This includes image data processing through binarization and sampling, as well as the application of the LBM for single-phase flow to compute the intrinsic permeability components.
Section~\ref{sec:Model1} presents the initial CNN model used in this study. It begins with a description of the model's architecture, data preparation for training, and hyper-parameter selection. This is followed by an analysis of the model's training results and its performance on unseen data.
The extension to an informed CNN model, achieved by incorporating physical parameters into the model's architecture, and its impact on model performance are discussed in Section~\ref{sec:Model2}. 
In Section~\ref{sec:Model3}, an alternative approach to improving the model's performance and generalization through data augmentation is discussed. Specifically, it begins with training an initial model on synthetic data generated via GAN to capture general patterns. This is followed by applying transfer learning, where the pre-trained model is fine-tuned on the original dataset to enhance its performance. Finally, concluding remarks are given in Section~\ref{sec:Conclusions}.
  

\section{Macroscopic saturated porous media model}
\label{sec:theoryContinum}
\subsection{Homogenization, volume fractions, and densities }
For the continuum mechanical description, the following formulations consider a saturated, two-phase porous material consisting of a materially incompressible solid but compressible solid matrix and a materially incompressible pore fluid.
The microscopically heterogeneous porous materials can be well described within the macroscopic TPM framework. Hereby, a homogenization process is applied to a representative elementary volume (REV), resulting in a smeared-out continuum $\varphi$ with overlapping, interacting, and statistically distributed solid and liquid aggregates $\varphi^\alpha$ ($\alpha=S$ for the solid phase and $\alpha=F$ for the pore-liquid phase), see \cite{Ehle02} for details.
Having immiscible solid and fluid aggregates, the volume fraction $n^\alpha:=\rmd v^\alpha/\rmd v$ of $\varphi^\alpha$ is defined as the ratio of the partial volume element $\rmd v^\alpha$ to the total volume element ${\rm d} v$ of $\varphi$. 
Furthermore, the saturation constraint of the fully saturated material is expressed as 
\begin{equation}
\Suma n^\alpha=n^S+n^F=1
\quad\mbox{with}\quad
\left\{\begin{array}{ll}
n^S&\mbox{:\,solidity}\,,
\\
n^F&\mbox{:\,porosity}\,.
\end{array}\right.
\label{eq:satconst}
\end{equation}
Two density functions are also defined, namely a material (effective or intrinsic) density $\RhoaR:=\rmd m^\alpha/\rmd v^\alpha$, and a partial density $\rho^{\alpha}:=\rmd m^\alpha/\rmd v$, with $\rmd m^\alpha$ being the local mass element of $\varphi^\alpha$. 
While $\RhoaR=\mathrm{const.}$ in the current treatment, i.e. materially incompressible solid and fluid constituents, the solid (bulk) matrix is still compressible through the change of $n^\alpha$. Hence, the partial and the effective densities are related by $\rho^{\alpha}:=n^\alpha\,\RhoaR$.

\subsection{Kinematics of multi-phase continua}
\label{sec:TheoKin}
Within the framework of continuum mechanics applied to multiphase porous media, the individual constituents $\varphi^\alpha$ are regarded to have unique states of motion, see, e.g., \cite{Bowe76, Haup93}.
Thus, with $\VX_\alpha$ as the position vector of $\varphi^\alpha$ with respect to the reference configuration and $\Vx$ as the position vector with respect to the current configuration, each constituent has an individual Lagrangian (material) motion function $\Vchi_\alpha$ and has its velocity field $\Vv_\alpha$, viz.
\begin{equation}
\label{eq:motion}
\arraycolsep 0.3ex
\begin{array}{l}
\Ds
  {\Vx}  = \Vchi_\alpha({\bf X}_\alpha,t)
  \ \,\Leftrightarrow\ \,
  {\VX}_\alpha = \Vchi_\alpha^{-1}({\Vx},t)
  \,
\quad
\mbox{and}
\quad
  {\Vv_\alpha}:= \Vxs_\alpha
  = \frac{\rm d_\alpha{\Vx}}{\rmd t}
\,
\end{array}
\end{equation}
with  $\Vchi_\alpha^{-1}$ being the inverse (Eulerian or spatial) motion function and
\begin{equation}
\arraycolsep 0.3ex
\begin{array}{l}
\Ds
\earg'_\alpha:=\frac{\rmd_\alpha\earg}{\rmd t}
=\frac{\p\earg}{\p t} + \grad\earg\cdot\Vv_\alpha 
\end{array}
\label{eq:dt}
\end{equation}
representing the material time derivative of an arbitrary vector quantity $\earg$ with respect to the motion of $\varphi^\alpha$ and $\grad\earg:={\p\earg}/{\p\Vx}$.
%
In this work, the motion of the solid phase is described using a Lagrangian approach via the solid displacement $\Vu_S$ and velocity $\Vv_S$. 
For the pore-fluid phase, the motion is described either by an Eulerian description using the fluid velocity $\Vv_F$ or by a modified Eulerian setting via the seepage velocity $\Vw_{F}$, i.e.
%
\begin{equation}
\label{eq:pfields}
  \Vu_S=\Vx-\VX_S
  \,,\,\,\,\,
  \Vv_S=(\Vu_S)'_S
  \,,\,\,\,\,
  \Vw_{F}=\Vv_F-\Vv_S
  \,.
\end{equation}
%
Within a  small strain assumption, the essential kinematic relation is the linearized small solid strain tensor $\Teps_S$, expressed as
\begin{equation}
\Teps_S\,=\,\Half\,(\grad\,\Vu_S\,+\,\grad^T\Vu_S)\,.
\label{eq:LinEps}
\end{equation}

%

\subsection{Local balance relations}

In the preparation of the governing balance relations that will be used for further discussion, we will make the following simplifying assumptions:
\begin{itemize}
  \setlength\itemsep{0.5em}
    \item[$\bullet$] Non-polar constituents with symmetric stress tensor ($\TT^\alpha=(\TT^\alpha)^T$). Thus, the angular momentum balance equation is always satisfied.
    \item[$\bullet$] A quasi-static biphasic model with negligible inertia terms, i.e.  $\rho^\alpha\,(\Vv_\alpha)'_\alpha\approx\Vnull$\,.
    %
    \item[$\bullet$] Isothermal conditions, which eliminates the need for the energy balance equation.
    \item[$\bullet$] No mass production $\hat{\rho}^\alpha$ between the solid and the fluid phases, i.e. $\hat{\rho}^\alpha \equiv 0$\,.
    %
\end{itemize}
%
%
Under these assumptions within the TPM, the constituent balance equations of the fluid-saturated porous media can be expressed as

\begin{itemize}
\itemsep -0.75ex
\item[$\bullet$] Constituent mass balance:
  \begin{equation}
  \label{eq:parvolbal}
 (\rho^\alpha)'_\alpha+\rho^\alpha\,\div\,\Vv_\alpha\,=\,0\,
  \end{equation}
\item[$\bullet$] Constituent momentum balance:
  \begin{equation}
  \label{eq:parmombal}
\Vnull=\,\div\,\TT^\alpha+\rho^\alpha\,\Vg+\Bhat{\Vp}^\alpha
  \end{equation}
\end{itemize}
%
In Eq.~(\ref{eq:parmombal}), $\Vg$ is the mass-specific gravitational force,
$\TT^\alpha$ is the symmetric partial Cauchy stress tensor, and $\Bhat{\Vp}^\alpha$ is the direct momentum production (local interaction force between $\varphi^S$ and $\varphi^F$). 
For this, the overall conservation of momentum results in $\Bhat{\Vp}^S+\Bhat{\Vp}^F=\Vnull$\,.
Considering  $\alpha=S$ in Eq.~(\ref{eq:parvolbal}) together with  $\RhoaR=\mathrm{const.}$ yields the solid volume balance equation. Applying an analytical integration to this (see, e.g., \cite{HeiderMarkert17, EhlersWagner2019}), we get the solidity as a secondary variable. In particular, we have for $\varphi^S$ 
\begin{equation}
\label{eq:solidity}
  (n^S)'_S+n^S\div\,\Vv_S=0
~~~\xrightarrow[]{\text{integration}}~~~
n^S= {\Ds n^S_{0}}\,\det\TF_S^{-1}
~~~\xrightarrow[]{\text{linearization}}~~~
n^S \approx {\Ds n^S_{0}}\,(1-\div\,\Vu_S)
\end{equation}
with $\TF_S=\p\Vx/\p\VX_S$ being the solid deformation gradient and $n^S_{0}$ the initial volume fraction of $\varphi^S$.
Therefore, the direct correlation between the solid deformation $\Vu_S$ and the porosity  ($n^F=1-n^S$) is apparent in Eqs.~(\ref{eq:solidity})$_{2,3}$.

%

\subsection{Effective stresses and permeability formulation}
\label{sec:TheoConstitutive}
According to the principle of effective stresses, the total stress state at any material point of the homogenized material consists of an `extra' or `effective' stress, denoted by the subscript $\earg_E$, and a weighted pore-fluid pressure term, see, e.g., \cite{Bish1959, dBEh90b,EHLERS201835_Effective_Stresses} for a review and references.
Thus, $\TT^\alpha$ and $\Bhat{\Vp}^\alpha$ with $\alpha=\{S,F\}$ in Eq.\,(\ref{eq:parmombal}) can be expressed as
\begin{equation}
\begin{array}{l}
\TT^S=\TT^S_E-n^S\,p\,\TI\,,\quad
\TT^F=\TT^F_E-n^F\,p\,\TI\,,\quad
\Bhat{\bf p}^F=\Bhat{\bf p}^F_E+p\,\grad\,n^F\,.
\end{array}
\label{eq:ExtraStresses}
\end{equation}
For a linear-elastic solid phase, the effective-stress tensor $\TT^S_E$ can be expressed as
\begin{equation}
\TT^S_E
\,=\,
\kappa^S\,\tr(\Teps_{S})\,\TI \,+\,2\,\mu^S\,\Teps^D_{S}
\label{eq:ElastStressDer}
\end{equation}
with $\kappa^S:=\lambda^S+\tfrac{2}{3}\mu^S$ being the bulk modulus of the porous solid matrix, which is defined in terms of the macroscopic \Lame\ constants $\mu^S$ and $\lambda^S$.
Moreover, $\tr(\Teps_{S}):= \Teps_{S}\cdot\TI=\div\,\Vu_S$ defines the scalar-valued trace of  the strain tensor $\Teps_{S}$
with $\Teps^D_{S}:=\Teps_{S}-\tfrac{1}{3}\tr(\Teps_{S})\TI$ as the deviatoric strain tensor.

In defining the fluid effective stress, we proceed with the assumptions of Newtonian fluid and negligible average normal viscous stress (i.e. Stokes' hypothesis). Thus, the effective or frictional fluid stress can be expressed as
\begin{equation}
\TT^F_E=\TT^F_\text{fric}=\,\mu^F(\grad\,\Vv_F+\grad^T\Vv_F)\,, 
\label{eq:TFE}
\end{equation}
with $\mu^F\,:=\,n^F\,\mu^{\FR}>0$ being the partial shear or dynamic viscosity.
Having possibly an anisotropic and deformation-dependent 2$^\text{nd}$-order intrinsic permeability tensor $\TK^S(\Vu_S)$, the constitutive equation of the effective or frictional momentum production  $\Bhat{\bf p}^F_E$  can be expressed as
\begin{equation}
\Bhat{\bf p}^F_E=\Bhat{\bf p}^F_\text{fric}=-\,{(n^F)\,\mu^{F}}{(\TK^S)^{-1}}\,\Vw_F\,.
\label{eq:PFE}
\end{equation}
A detailed derivation and discussion of the above constitutive relations based on the 2$^\text{nd}$-law of thermodynamics can be found in \cite{Ehle02, Ehlers2022_DarcyForchheimerBrinkman}.


\subsection{Governing balance equations}
Considering the constituent balance relations (\ref{eq:parvolbal}, \ref{eq:parmombal}) together with the relations between the total and effective stresses in (\ref {eq:ExtraStresses}), the governing equations of the binary model to determine $\{\mathbf{u}_S,\,\mathbf{v}_F,\,p\}$ can be expressed as follows:
\begin{equation}
\begin{array}{lcl}
\bullet\,\,\, \text{Solid momentum balance:} & &
\Vnull \,=\, \text{div}\big(\TT^S_E\big) - n^S\, \text{grad}\,p  + \rho^S\Vg - \hat{\Vp}^F_E \\[3mm]
\bullet\,\,\, \text{Fluid momentum balance:} & &
\Vnull \,=\, \text{div}\big(\TT^F_E\big) - n^F\, \text{grad}\,p + \rho^F\Vg + \hat{\Vp}^F_E \\[3mm]
\bullet\,\,\, \text{Overall volume balance:} & &
0 \,=\, \text{div}\big(\Vv_S \,+\, n^F\,\Vw_F\big) \\
\end{array}
\label{eq:GovEqu}
\end{equation}
Therein, (\ref{eq:GovEqu})$_{1,2}$ represent the momentum balance equations of the individual constituents, i.e. solid and fluid, with corresponding momentum interaction terms $\hat{\Vp}^S_E=-\hat{\Vp}^F_E$.
Equation (\ref{eq:GovEqu})$_3$ represents the overall volume balance as the sum of the solid and fluid volume balances. 
In this context, it is worth mentioning that the set of equations~(\ref{eq:GovEqu}) to describe the hydromechanical response of the biphasic model is not unique as discussed in \cite{Maea10, Ehlers2022_DarcyForchheimerBrinkman}. Several alternative multi-field formulations can be presented, such as using the overall momentum balance instead of the solid momentum balance  or merging the fluid momentum balance with the overall volume balance. These reformulations can significantly impact the numerical stability and the way the boundary conditions are formulated as discussed in, e.g., \cite{Maea10}.

\subsection{Porous media flow models}
\label{sec:PMFlow}

The three equations in (\ref{eq:GovEqu}) describe pore-fluid flow through homogenized and deformable porous media, where 
$n^S$ in a linear solid model is a function of $\Vu_S$, i.e. it varies during deformation according to Eq.~(\ref{eq:solidity})$_3$\,. However, in the pore-scale flow models using the LBM, the simulation is applied to a micro-geometry at a fixed state of deformation (corresponding to a specific constant value of $n^S$), i.e. no solid deformation occurs during the individual LBM simulation. 
This allows for a connection to the classical macroscopic Darcy and Brinkman equations, which were originally developed under the assumption of a non-deforming solid skeleton (i.e., $\Vu_S=\Vnull$ and $n^S=\text{const.}$), making them a special case of the broader TPM model. 
In the following sections, the Darcy and Brinkman flow models will be briefly discussed in relation to the fundamentals of TPM for two-phase porous materials, following the pioneering work of \citet{Ehlers2022_DarcyForchheimerBrinkman}.


\subsubsection{Darcy flow}
Darcy's law is widely used as a constitutive assumption for the description of mostly fully saturated pore flow conditions in porous media. 
It is primarily applied to steady-state, incompressible, laminar flow in media with small pore spaces, where the drag force dominates over the frictional force. 
Thus, to recover a Darcy-like flow model, the frictional force (related to $\TT^F_E$) is neglected from the fluid momentum balance (\ref{eq:GovEqu})$_2$ in comparison with the drag force (related to $\hat{\Vp}^F_E$). This neglect can be mathematically justified following a dimensional analysis, as discussed in detail in \cite{Ehlers2022_DarcyForchheimerBrinkman}. In particular,
having $\text{div}\big(\TT^F_E\big)\ll\hat{\Vp}^F_E$
yields $\text{div}\big(\TT^F_E\big)\approx\Vnull$\,.
Thus, the fluid momentum balance (\ref{eq:GovEqu})$_2$ yields after neglecting $\text{div}\big(\TT^F_E\big)$ and considering the definition of the effective momentum production (\ref{eq:PFE}) the Darcy-like flow model, expressed as 
%
\begin{equation}
    \text{grad}\,p=- \mu^F\,\Big(\TK^S \Big)^{-1}\Vw_F+\rho^\FR\,\Vg\,,
\label{eq:DarcyEqGrav}
\end{equation}
 which accounts for both pressure-driven and gravity-driven flow.
Although the gravitational force ($\rho^\FR\Vg$) can play a significant role in some fluid mechanics applications, they are often negligible in small-scale systems where pressure gradient effects dominate. In our study, this holds true when applying Darcy's law in an inverse technique to estimate the permeability parameter based on data from LBM simulations conducted on mesoscale RVEs (cf. section~\ref{InrinsicPermComputation}). Consequently, an additional simplification can be applied to Eq.~(\ref{eq:DarcyEqGrav}), resulting in a simple flow model
\begin{equation}
    \text{grad}\,p=- \mu^F\,\Big(\TK^S \Big)^{-1}\Vw_F\,.
\label{eq:DarcyEq}
\end{equation}
In analogy to Darcy's law, this equation establishes that the flow velocity of a fluid through a porous medium is linearly proportional to the pressure gradient driving the flow, with the intrinsic permeability $\TK^S$ and the fluid dynamic viscosity $\mu^F$ acting as the proportionality constants.
Using  $\TK^S$ in the above formulation provides a more robust and fluid-independent measure of a porous medium's ability to transmit fluids. Since $\TK^S$ depends primarily on the micro-morphology of the porous medium, it can be accurately derived from micro-CT images of the material, as will be discussed in section~\ref{InrinsicPermComputation}.

%
\subsubsection{Darcy–Brinkman  flow}
In this work, the database for the ML model is generated by applying the LBM to simulate fluid flow on the microscale. The boundary conditions of the corresponding 3D pore-scale samples are designed to maintain laminar flow at a low Reynolds number. Consequently, Darcy's law is applied to inversely determine the intrinsic permeability.
On the macroscopic scale, incorporating the Brinkman model allows the approach to capture viscous forces when necessary by introducing an additional effective fluid stress term. This ensures that the model can account for both permeability and viscous effects in cases where they become significant.

The Brinkman or Darcy–Brinkman flow model extends the Darcy flow model by incorporating the fluid viscous shear stress. 
This extension allows for a more accurate representation of heterogeneous media and improves flow modeling near boundaries and in transitional flow regimes.
To account for both the drag force, exerted by the porous matrix, and the viscous shear stress of the pore fluid, \citet{Brinkman1949} suggested combining Darcy's law with the Navier–Stokes equation.
A similar result can be realized within the TPM by not neglecting the term $\text{div}\big(\TT^F_E\big)$ in the fluid momentum balance equation. Thus, considering the definition of the effective fluid stress (\ref{eq:TFE}) in the fluid momentum balance (\ref{eq:GovEqu})$_2$ and assuming creeping flow conditions with $(\Vv_F)'_F\approx\Vnull$ yields
\begin{equation}
\text{grad}\,p=- \mu^F\,\Big(\TK^S \Big)^{-1}\Vw_F+\mu^\FR\,\Delta\Vv_F
+ \rho^\FR\Vg
\label{eq:DarcyBrinkmanEq}
\end{equation}
which is considered a refined version of the Brinkman or Darcy–Brinkman flow equation.
In this, $\Delta\Vv_F$ represents the Laplacian of the velocity field, which accounts for the viscous shear stresses under the assumption of incompressible flow.
As in Darcy's flow (\ref{eq:DarcyEq}), having $\TK^S$ in the above formulation provides a fluid-independent measure of a porous medium's ability to transmit fluids. Thus, deriving $\TK^S$ from micro-CT images together with knowing the fluid properties allows for the modeling of nonlinear flow within porous media.
%
%
 
Other models exist in the literature that describe flow through porous media, including the Forchheimer or Darcy–Forchheimer model, which extends Darcy's law by incorporating nonlinear effects that become significant at higher flow velocities. Specifically, The Forchheimer equation adds terms to account for these nonlinearities, which depend on factors such as tortuosity and velocity (see, e.g., \cite{markert2007constitutive, Ehlers2022_DarcyForchheimerBrinkman}). While these models are important for accurately capturing flow behavior under specific conditions, a detailed exploration of them is beyond the scope of this work, which focuses primarily on determining intrinsic permeability.

\section{Database generation}
\label{sec:Database}

The database utilized in this study is derived from $\mu$-CT images of Bentheim sandstones, accessible via the Digital Rock Portal \cite{datasetBentheimer} and hold a resolution of $\delta x_1=\delta x_2=\delta x_3=8.96 \, \mu \mathrm{m}$ per pixel. These images are obtained at various deformation states, which correspond to volumetric strains $\varepsilon^V\!\in \{2\%,\, 4\%,\, 6\%,\, 8\%,\, 10\%,\, 20\%,\, 30\%,\, 40\%\}$. Each deformation state comprises 8 samples, each consisting of 500-700 slices of 700$\times$700 pixels.
 
\subsection{Image processing}
\label{sec:ImageProcessing}

\begin{figure}[!ht]
\begin{center}
\includegraphics[width=8.5cm]{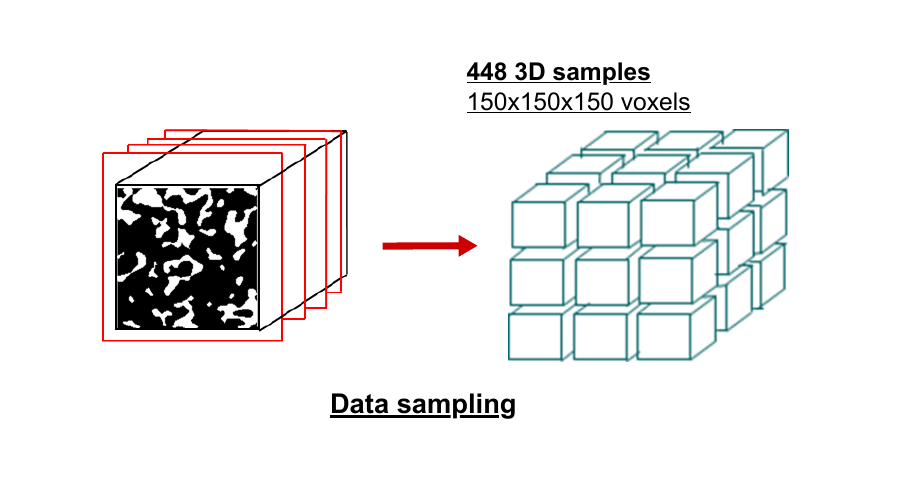}
\includegraphics[width=7.0cm]{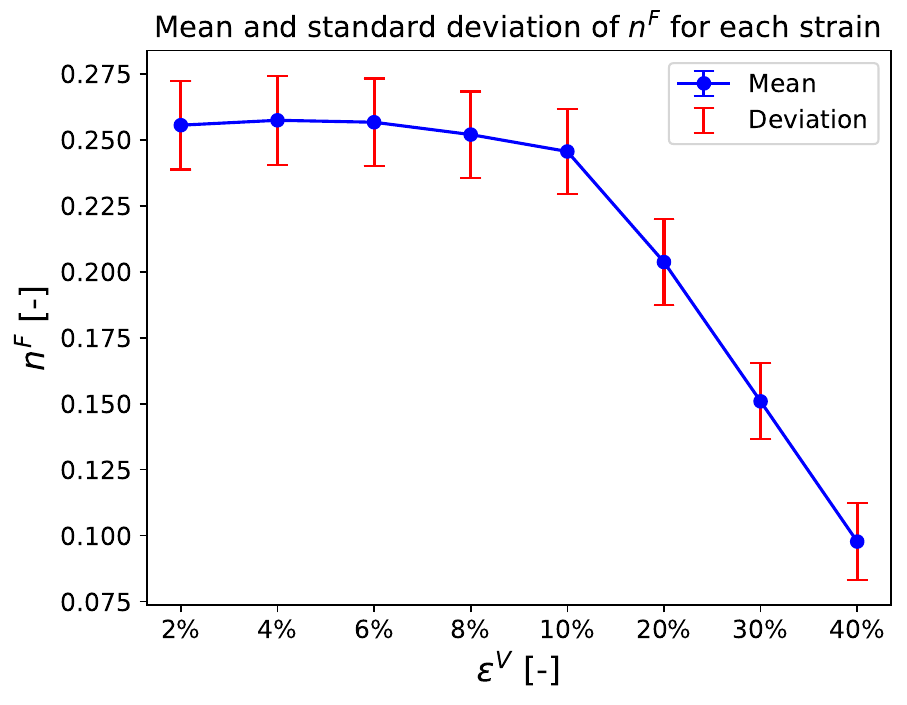}
\end{center}
\caption{Illustration of data sampling (left). 
Mean and standard deviation of the porosity $n^F$ for each strain level $\varepsilon^V$ showing the deformation-dependency, while the 448 samples are considered (right).}
\label{Fig:Sampling_448_porosity}
\end{figure}

In database preparation for the training, the raw CT images of the 8  three-dimensional samples of 700$\times$700$\times$700 voxels undergo binarization and sampling. This process results in a sufficient database size of 448 three-dimensional (3D) samples. 
With this, each of the downscaled 3D samples has a size of 150$\times$150$\times$150 voxels to ensure manageable data size while preserving essential microstructural features. This treatment is consistent with the work of \citet{HongLiu2020_CNN_Permeability}. An illustration of data sampling to get 448 3D samples is presented in Fig.~\ref{Fig:Sampling_448_porosity}, left.

Moreover, the binarization step is performed using thresholding to convert grayscale CT images into binary images, where the pore space and the solid matrix are distinguished. This allows the porosity ($n^F$) of the material to be estimated for each strain level, showing the deformation-dependency as illustrated in Fig.~\ref{Fig:Sampling_448_porosity}, right.
Based on these binary images, the specific surface area (SSA), which is the ratio of the total surface area of the material per unit volume, can also be computed with the help of a Matlab code published by \citet{degruyter2010synchrotron}.
In addition, these binary images serve as the basis for simulations performed in Palabos \cite{Palabos2020}, an open-source framework for lattice Boltzmann simulations.

\subsection{LBM for single-phase fluid flow}
\label{OneFluid_LBMTheor}
In this study, we employ lattice Boltzmann method (LBM) simulations to calculate the average fluid velocity within each 3D sample under an applied pressure gradient.
This allows for the inverse calculation of the intrinsic permeability tensor using the Darcy flow equation presented in (\ref{eq:DarcyEq}) with $\Vv_S=\Vnull$.
Abstract introduction to the LBM in the context of single fluid flow is given below, while more details and references can be found in \cite{chaaban2020upscaling,chaaban2022_TwoPhaseLBMtpm,ChaabanPAMM2022}.

The LBM uses a mesh-based approach to solve the Boltzmann equation \cite{boltzmann2012lectures}.
It starts by defining the velocity distribution function $f(\mathbf{x}, {\Vxi}, t)$, which represents the probability of finding a fluid particle at a specific position $\Vx$ and time $t$ with a certain discrete velocity 
$\Vxi$\,. The Boltzmann equation then describes how $f(\mathbf{x}, {\Vxi}, t)$ evolves over space and time. 
The evolution of $f(\mathbf{x}, {\Vxi}, t)$, associated with the exchange of momentum and energy amongst these particles, occurs through two key processes: streaming and collision, viz.
\begin{equation} 
\label{Boltzmann}
\dfrac{df}{dt} \Big|_{\text{streaming}} = \dfrac{df}{dt} \Big|_{\text{collision}}\,\quad\text{with}\quad\quad
\underbrace{\dfrac{\partial f}{\partial t} + 
\Vxi \cdot \dfrac{\partial f}{\partial \mathbf{x}}}_{\text{streaming operator}} = \underbrace{\vphantom {\dfrac{\partial f}{\partial t} + \mathbf{v} \cdot \dfrac{\partial f}{\partial \mathbf{x}}} \Omega \, (f)\,.}_{\text{collision operator}}
\end{equation}
As described in \citet{kruger2017lattice} and \citet{HeLuo_LBM_1997}, the distribution function $f(\mathbf{x}, {\Vxi}, t)$ is linked to macroscopic variables such as fluid density $\rho^\FR$ and fluid velocity $\Vv_F$ through its moments. This connection is established using the following integrals:
%
\begin{equation}
\begin{array}{l}
\Ds
\rho^\FR(\mathbf{x},t)\approx\rho_l\,(\mathbf{x},t) = \int{f(\mathbf{x},\Vxi,t)} \, d\Vxi \quad \text{and} \quad
\Vv_F(\mathbf{x},t)\approx\Vu_l\,(\mathbf{x},t) = \dfrac{1}{\rho_l}\int{\Vxi\,f(\mathbf{x},\Vxi,t)} \, d\Vxi\,.
\end{array}
\end{equation}

%
For the spatial discretization in 3D, a fluid particle is restricted to stream in 19 possible directions, known as D3Q19. These directions are defined as:
\begin{eqnarray}
\mathbf{e}_i = \left\{\begin{array}{lcl}
(0,0,0) & \quad & i=0 \\
(\pm1,0,0), (0,\pm1,0), (0,0,\pm1) & \quad & i=1, 2,\dotsc, 6\\
(\pm1,\pm1,0), (\pm1,0,\pm1), (0,\pm1,\pm1) & \quad & i=7, 8,\dotsc, 18, \\
\end{array}\right.
\end{eqnarray}
where $\mathbf{e}_i$ is the direction of the velocity vectors $\Vxi_i = c \, \mathbf{e}_i\,$ given in terms of $c$ as the ratio of the distance between the nodes $\Delta x$ to the time-step size $\Delta \,t$\,.
Regarding the collision operator $\Omega \, (f)$ in (\ref{Boltzmann}), the Bhatnagar-Gross-Krook (BGK) \cite{PhysRev.94.511} model is used since it is easy to implement and has been widely used in LBM fluid flow simulation \cite{wolf2004lattice}. 
In particular, the BGK collision operator $\Omega_{\text{BGK}}$ is expressed as
\begin{eqnarray}
\label{eq_BGK}
\Omega_{\text{BGK}} = - \frac{f_i - f_i^{eq}}{\tau}  \quad \text{with} \quad \tau := \frac{1}{2} + \nu_l \, c_s^{-2}\,.
\end{eqnarray}
Herein, the relaxation time $\tau$ depends on the lattice fluid viscosity $\nu_l$ and lattice speed of sound $c_s=1/\sqrt{3}$\,. 
The BGK model facilitates the relaxation of the distribution functions $f_i$ toward equilibrium distributions $f_i^{eq}$ at a collision frequency $\tau^{-1}$. The formulation of $f_i^{eq}$ is expressed as follows
\begin{eqnarray}
\label{eq_equil}
f_i^{eq} = w_i \, \rho_l \, \bigg(1 + \dfrac{\mathbf{e}_i \cdot \mathbf{u}_l}{{c_s}^2} + \dfrac{(\mathbf{e}_i \cdot \mathbf{u}_l)^2}{2{c_s}^4} - \dfrac{(\mathbf{u}_l \cdot \mathbf{u}_l)^2}{2{c_s}^2}\bigg)\,,
\end{eqnarray}
where $w_i$ presents the lattice weights, i.e.
\begin{eqnarray}
w_i = \left\{\begin{array}{lcl}
1/3 & \quad & i=0 \\
1/18 & \quad & i=1, 2,\dotsc, 6 \\
1/36 & \quad & i=7, 8,\dotsc, 18\,.
\end{array}\right.
\end{eqnarray}
The distribution functions are updated through the following equation:
\begin{eqnarray}
\underbrace{f_i(\mathbf{x} + \Vxi_i \Delta t, t + \Delta t) - f_i(\mathbf{x}, t)}_{\text{streaming}} = \underbrace{\Omega_{\text{BGK}}}_{\text{collision}}.
\end{eqnarray}
For the boundary conditions (BCs), the Zou-He \cite{zou1997pressure} bounce-back boundary dynamics are used. For a more detailed explanation of the LBM approach for single-phase fluid flow, refer to \cite{chaaban2020upscaling}.


\subsection{Intrinsic permeability computation}
\label{InrinsicPermComputation}

\begin{figure}[!ht]
\begin{center}
\includegraphics[width=7.2cm]{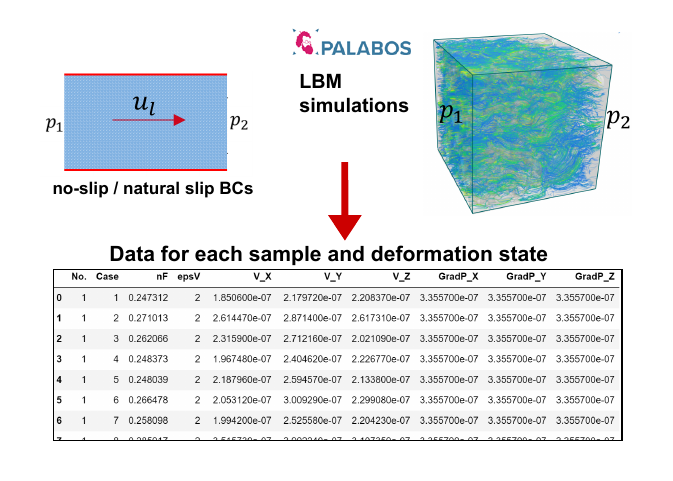}
\quad\quad
\includegraphics[width=7.2cm]{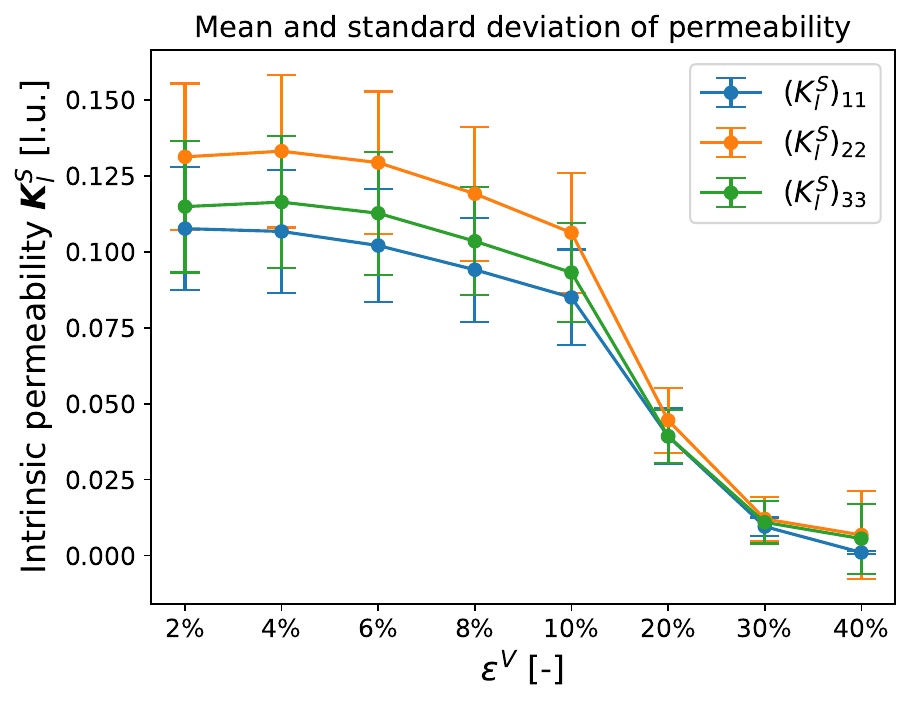}
\end{center}
\caption{Illustration of data generation using LBM with prescribed pressure difference and no-slip/natural slip BCs (left). 
Mean and standard deviation of the intrinsic permeability components $K^S_{ii}$ for each strain level $\varepsilon^V$ showing the deformation-dependency and anisotropy (right).}
\label{Fig:LBM_illustraion_KS_Diag}
\end{figure}

The proposed CNN model in this work uses CT images of sandstone as input and outputs the components of the permeability tensor. 
To consider the intrinsic permeability components in the database, 3D samples of Bentheim sandstone at varying levels of deformation are input into the single-phase LBM solver described in section~\ref{OneFluid_LBMTheor}. 
The primary goal of the LBM simulation is to calculate the average lattice fluid velocity for each prescribed pressure gradient applied across the porous domain along the hydrodynamic axes $\mathbf{x}_1,\,\mathbf{x}_2$ and $\mathbf{x}_3\,$, i.e. $\nabla p_1, \,\nabla p_2 \, \text{and} \, \nabla p_3\,$, respectively. 
The simulation results are used to determine the lattice intrinsic permeability tensor
$\mathbf{K}^S_l$ in lattice units [l.u.]. This is done under the assumption that the permeability tensor is symmetric and positive definite \cite{kuhn2015stress}. 
As proposed by \citet{kuhn2015stress}, two fluid flow simulations for each direction with different boundary conditions are to be carried out: 
One with no-slip boundary conditions and another with natural slip boundary conditions on surfaces parallel to the fluid flow direction, see Fig.~\ref{Fig:LBM_illustraion_KS_Diag}, left, for illustration. The rationale behind this is that the average velocity computed in the pressure direction is lower with no-slip boundary conditions compared to that with natural slip boundary conditions. The difference between these velocities indicates additional fluid flow in the direction orthogonal to the pressure gradient. 
The average velocities for the no-slip boundary conditions are then used to compute the diagonal components of the permeability tensor using a Darcy filter law:
%
\begin{equation}
\label{eqnDarcy}
(K_l^S)_{ii}  = 
- \nu_l \, \frac{( u_{l})_{i,\,\text{avg}}}{\nabla\!_i \, p_l}\quad\big[\mathrm{l.u.}\big]\,, \quad \mathrm{with} \quad i=1,\,2,\,3.
\end{equation}
In this, the lattice pressure gradient, presented by $\nabla\!_i \, p_l = \partial p_l / \partial x_i$\, is induced between two opposing surfaces perpendicular to the flow direction to calculate the average lattice fluid velocity $(u_{l})_{i,\,\text{avg}}$\,. 
As for the latter, the unknown non-diagonal elements of the permeability tensor are computed using the average velocity with natural slip boundary conditions following \cite{kuhn2015stress} as
\begin{eqnarray} 
\label{eqnNonDiag}
\begin{bmatrix}
	\nabla\!_2 \, p_l & \nabla\!_3 \, p_l & 0 \\[1mm]
	\nabla\!_1 \, p_l & 0 & \nabla\!_3 \, p_l \\[1mm]
	0 & \nabla\!_1 \, p_l & \nabla\!_2 \, p_l 
\end{bmatrix}
\begin{bmatrix}
(K_l^S)_{12} \\[1mm] (K_l^S)_{13} \\[1mm] (K_l^S)_{23} 
\end{bmatrix}
=
\begin{bmatrix}
- \nu_l \, ( u_{l})_{1,\,\text{avg}} - (K_l^S)_{11} \, \nabla\!_1 \, p_l \\[1mm] 
- \nu_l \, ( u_{l})_{2,\,\text{avg}} - (K_l^S)_{22} \, \nabla\!_2 \, p_l \\[1mm] 
- \nu_l \, ( u_{l})_{3,\,\text{avg}} - (K_l^S)_{33} \, \nabla\!_3 \, p_l
\end{bmatrix}.
\end{eqnarray}
The non-diagonal components computed for Bentheim sandstone are much smaller than the diagonal components. Thus, we neglect them for simplicity from the ML model.
In particular, the symmetric permeability tensor and its simplified diagonal form are expressed as follows
\begin{equation}
    \TK_l^S =
    \begin{bmatrix}
	(K_l^S)_{11} \;& (K_l^S)_{12} \;& (K_l^S)_{13} \\[1mm]
	(K_l^S)_{21} \;& (K_l^S)_{22} \;& (K_l^S)_{23} \\[1mm]
	(K_l^S)_{31} \;& (K_l^S)_{32} \;&(K_l^S)_{33}
    \end{bmatrix}(\overline{\Ve}_i\otimes\overline{\Ve}_j)
    \approx
    \begin{bmatrix}
	(K_l^S)_{11} & 0 & 0 \\[1mm]
	0 & (K_l^S)_{22} & 0 \\[1mm]
	0 & 0 &(K_l^S)_{33}
    \end{bmatrix}(\overline{\Ve}_i\otimes\overline{\Ve}_j)\,.
\end{equation}
Here, $\overline{\Ve}_i$, $\overline{\Ve}_j$ represent the cartesian basis vectors with $i ,j\in\{1,2,3\}$ and $\otimes$ is the dyadic product (tensor product).
In this connection, Fig.~\ref{Fig:LBM_illustraion_KS_Diag}, right, shows the mean and standard deviation of the diagonal intrinsic permeability components $(K^S_l)_{11},\,(K_l^S)_{22},\,(K_l^S)_{33}$ for each strain level $\varepsilon^V$ showing the deformation-dependency and anisotropy in the flow.
The macroscopic intrinsic permeability tensor is derived from the lattice permeability tensor as follows
\begin{equation}\label{eqnConversion}
K_{ij}^S = (K_l^S)_{ij} \, (\delta x_i\,\delta x_j) \quad\text{in}\quad \big[\mathrm{m}^2\big] \,,
\end{equation}
where $\delta x_i$ and $\delta x_j$ characterize the spatial resolution of the $\bm{\mu}$-CT images in the $i$ and $j$ directions, respectively. 

For simplicity, we will refer to the permeability tensor as $(\TK^S)$, regardless of whether it is in [l.u.] or [$\mathrm{m}^2$], as this is only a unit change and does not affect the basic idea of the paper or the machine learning algorithms.




\section{Model~(1): $[K^S\!-\!n^F]$-CNN model}
\label{sec:Model1}
Convolutional Neural Networks (CNN), as a class of deep learning Artificial Neural Networks (ANN), is widely used in applications such as image classification, object detection, and text recognition. It mostly requires data that have a grid-like nature, such as images or time series data, see for example \cite{GU2018CNN, EIDEL2023_CNN, Aldakheel_2023_CNN_Magnetostatics,Dhillon_Verma_2020_reviewCNN,Aldakheel_2023_CNN_Hetero, Tandale_2023_CNN_RNN, BisharaEtAl2023_Review_ML_Multiscale} for review and applications.
CNN is considered as a sort of Feed-Forward NN (FFNN) approach with weight-sharing properties achieved through convolutional and pooling (subsampling) layers.
In this work, the CNN models are implemented using Python~3.11.7, utilizing the deep learning open-source code Keras \cite{chollet2015keras} with TensorFlow~2.15.0 \cite{abadi2016tensorflow} as the backend engine. The implementation and training processes are conducted in a Jupyter Notebook within an Anaconda environment. The training is performed on a NVIDIA~H100 GPU with 80~GB of memory.

The following discussion presents a brief explanation of the proposed CNN model architecture, which considers $\mu$-CT images of Bentheim sandstones as input to predict the associated permeability components. Additionally, the porosity is included as part of the output, enabling the model to consider the deformed state of the material. The discussion will also cover the model's training process and its performance on unseen data.

\subsection{CNN-model architecture and hyper-parameters}
\label{sec:Mode1Arch}

\begin{figure}[!ht]
\begin{center}
\includegraphics[width=12.0cm]{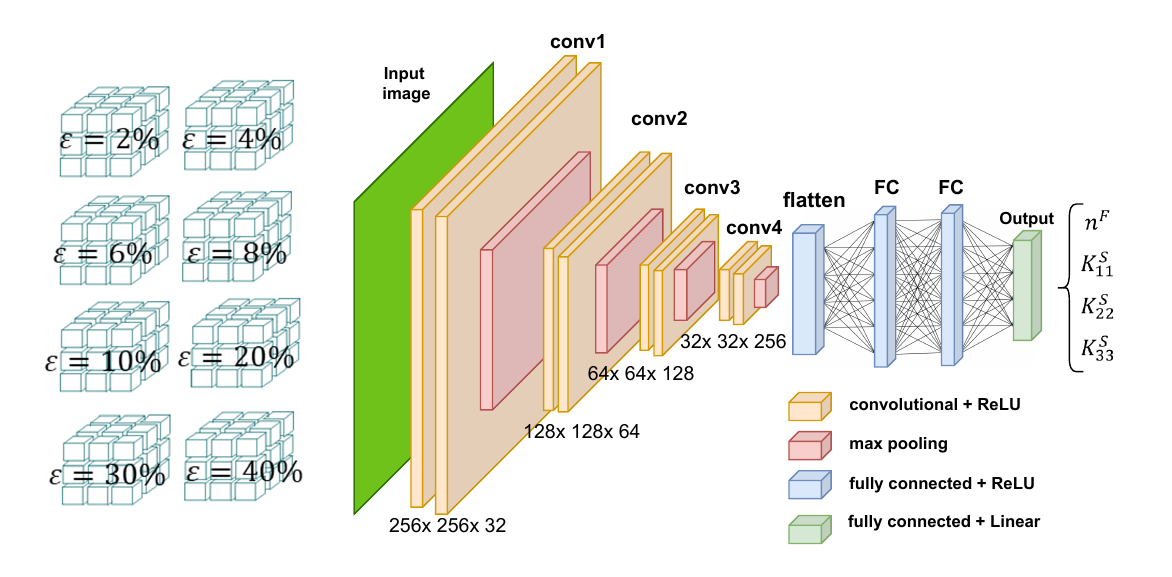}
\end{center}
\caption{Architecture of the 3D CNN illustrating the flow of information from the input image through multiple convolutional layers, max pooling layers, and fully connected layers. The input image is a 3D volume with dimensions \(256 \times 256 \times 32\), which passes through four convolutional layers (Conv1 to Conv4), each followed by max pooling to reduce spatial dimensions. After flattening, the feature maps are processed by two fully connected (FC) layers before generating the final output predictions (\(n^F\), \(K^S_{11}\), \(K^S_{22}\), \(K^S_{33}\)). }
\label{Fig:CNN_StrucIni}
\end{figure}

The CNN model comprises several key components, which are described as follows:
\begin{itemize}
    \item[$\bullet$] {\it Convolution Operation}:
     This step involves the application of a 2D filter (also known as a kernel or feature detector) to the input data, resulting in the generation of a feature map. This process highlights important features within the CT images by detecting patterns and edges that are crucial for permeability and porosity predictions.
    \item[$\bullet$] {\it Pooling Operation}: After the convolution operation, a pooling layer is applied to the feature map. This process aims to down-sample the feature map by calculating, for instance, their maximum or average values and sending only the significant features to the next  CNN layer.
    \item[$\bullet$] {\it Multi-Layer Perceptron (MLP)}:  The MLP consists of fully-connected (dense) layers that take the pooled feature map as input and produce a 1D feature vector as output.
\end{itemize}
The architecture of the proposed CNN model is illustrated in Fig.~\ref{Fig:CNN_StrucIni} and it is in line with the model presented in \cite{Aldakheel_2023_CNN_Magnetostatics}.

\noindent
The following details are associated with the underlying model:
\begin{itemize}
    \item[$\bullet$] \underline{Data preparation:}
    \begin{itemize}[label=$\circ$]
        \item The image-related input data is transposed and reshaped to ensure compatibility with the 3D CNN model architecture.
        \item Before splitting into training, validation, and test subsets, data indices are shuffled to randomize the samples.
        \item After Splitting, the output data ($n^F$ and $K^{S}_{kk}$) is scaled using the MinMaxScaler, class of the sklearn.preprocessing toolkit \cite{pedregosa2011scikit}, to normalize the values between 0 and 1, which is essential for stabilizing the training process.
    \end{itemize}
    \item[$\bullet$]  \underline{Model architecture:}
    \begin{itemize}[label=$\circ$]
        \item The CNN model features four convolutional blocks, each comprising two 3D convolutional layers with increasing filter sizes (32, 64, 128, 256) and a kernel size progressively increasing from 3$\times$3$\times$3 to 7$\times$7$\times$7. 
        \item Each convolutional layer employs  \textit{ReLU} activation and same padding to maintain spatial dimensions. 
        \item Following each convolutional block, a MaxPooling3D layer with a pool size of 2$\times$2$\times$2 is incorporated to down-sample the spatial dimensions, reducing computational load and mitigating overfitting.
    \end{itemize}
    %
    %
    %
    \item[$\bullet$]  \underline{Fully-connected (dense) layers:}
    \begin{itemize}[label=$\circ$]
        \item Post convolutional and pooling operations, a Flatten layer transforms the 3D feature maps into a 1D feature vector.
        \item This 1D vector is then fed into a series of fully connected (Dense) layers, configured with 64 and 32 units respectively, both employing  \textit{ReLU} activation and $L_2$ regularization to prevent overfitting. 
        \item The final Dense layer contains 4 units (related to $n^F$ and $K^{S}_{kk}$ with $k=1,2,3$) with a ``linear'' activation function, suitable for regression tasks.
    \end{itemize}
    \item[$\bullet$]  \underline{Loss Function and Optimization:}
    \begin{itemize}[label=$\circ$]
        \item The model is compiled using the Mean Squared Error (MSE) loss function, i.e.
        \begin{equation}\label{eq:LossMSE}
            {\cal D}_\text{MSE}=\frac{1}{n}\sum_{i=1}^n\Big[\big(n^{F,p}_{i}-n^{F,t}_{i}\big)^2+\big(K^{S,p}_{kk,i}-K^{S,t}_{kk,i}\big)^2\Big]\,,    
        \end{equation}
        where $n$ is the number of output data points and $k=1,2,3$\,.
        \item Optimization is managed by the Adam optimizer, configured with a learning rate of $0.00001$\,.
    \end{itemize}
    %
    %
     %
    %
    %
    \item[$\bullet$]  \underline{Training and Callbacks:}
    \begin{itemize}[label=$\circ$]
        \item The training process spans up to 500 epochs, i.e 500 complete passes through the entire training dataset, with a batch size of 16, comprising validation data to monitor performance. 
        \item Callback mechanisms such as \textit{ReduceLROnPlateau}, \textit{ModelCheckpoint}, and \textit{EarlyStopping} are implemented. 
    \end{itemize}
\end{itemize}

\subsection{Training and testing of Model~(1)}
\label{sec:TrainingResultsModel1}

\begin{figure}[!ht]
\begin{center}
\includegraphics[width=7.5cm]{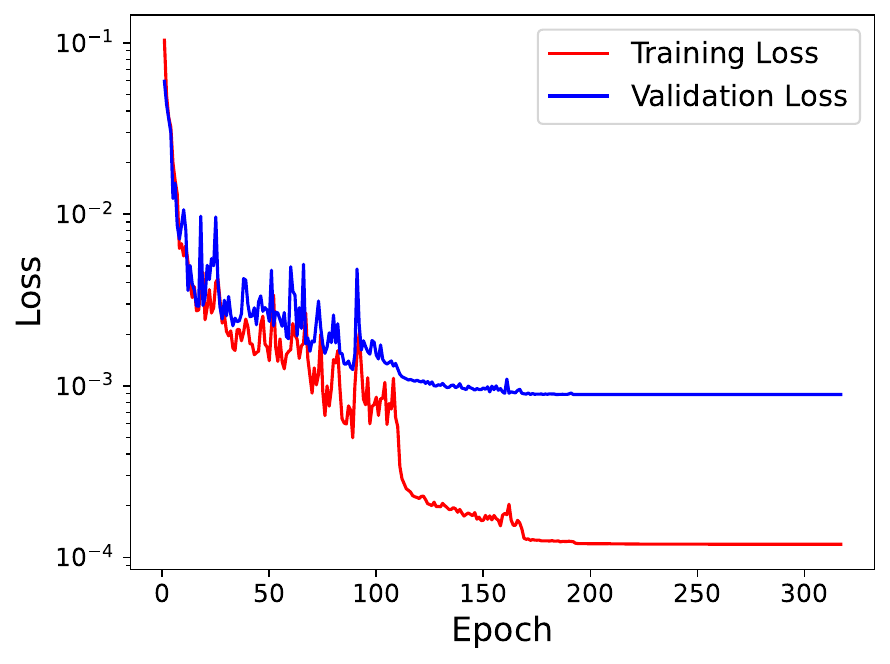}
\end{center}
\caption{Model~(1): Loss function value over the number of weight updates (Epoch).}
\label{Fig:lossEpocModel1}
\end{figure}

The training process is monitored by evaluating both the training and validation losses over epochs, with the corresponding curves shown in Fig.~\ref{Fig:lossEpocModel1}. Initially, both losses decrease rapidly, indicating that the model is learning effectively. As training continues, the training loss stabilizes at a lower value compared to the validation loss, suggesting that the model is encountering some degree of overfitting.
Changing the different hyperparameters of the CNN model for the given dataset did not help to eliminate the overfitting. Therefore, the state of the model after about 100 epochs (Loss $\approx10^{-3}$) is considered optimal and used for testing on unseen data.

\begin{figure}[!ht]
\begin{center}
\includegraphics[width=7.0cm]{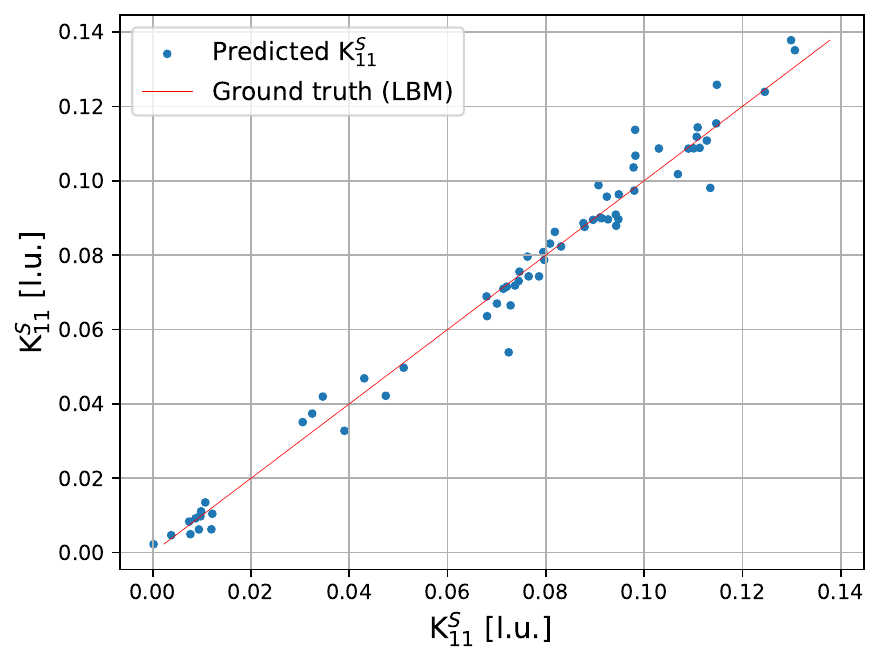}
\includegraphics[width=7.0cm]{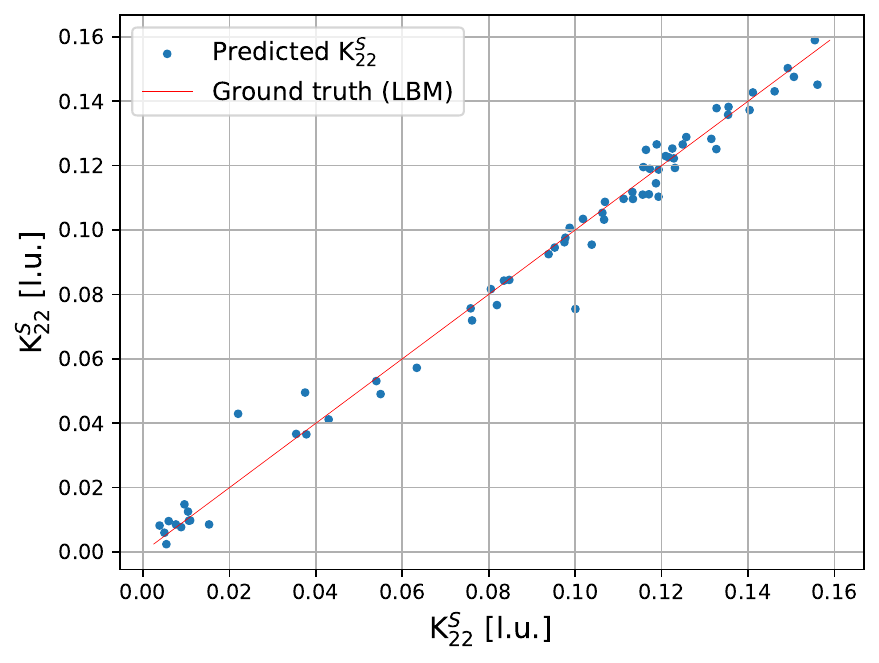}
\includegraphics[width=7.0cm]{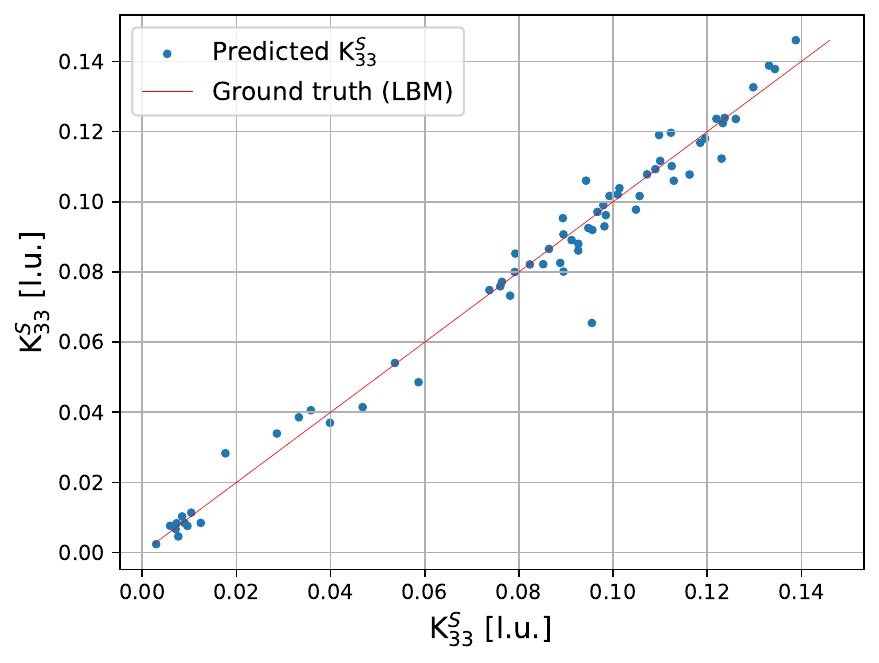}
\includegraphics[width=7.0cm]{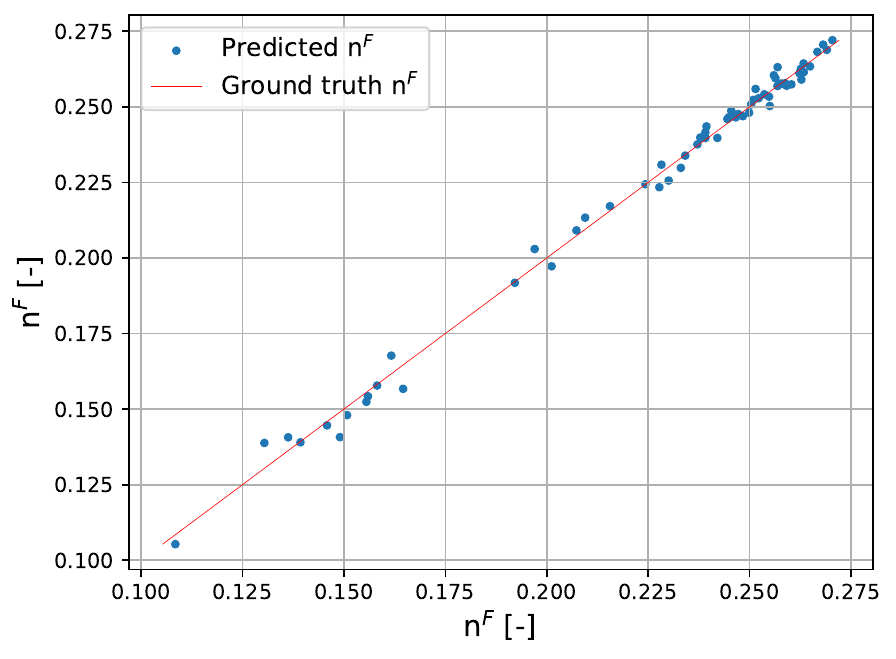}
\end{center}
\caption{Model~(1): CNN-predictions vs. ground-truth values of the intrinsic permeability components and porosity.}
\label{Fig:Model1TruePred}
\end{figure}

The scatter plots in Fig.~\ref{Fig:Model1TruePred}, which compare the predicted values to the ground truth values for the output variables \(n^F\) and \(K^{S}_{kk}\) (for \(k=1,2,3\)), further demonstrate the ability of the model to accurately predict the intrinsic permeability values and the corresponding porosity.
The model's performance can quantitatively be assessed using the \(R^2\)-score%
\footnote{
The \(R^2\) score, typically between 0 and 1, is calculated using the formula:
\begin{eqnarray}\label{Eq:R2}
R^2 = 1 - \frac{\sum_{i=1}^n (y_i - h_i)^2}{\sum_{i=1}^n (y_i - \bar{y})^2}
\quad\text{with}\quad
\left\{\begin{array}{ll}
y_i & \text{:True value for the \(i\)-th data point}, \\
h_i & \text{:Predicted value for the \(i\)-th data point}, \\
\bar{y} & \text{:Mean of the true values}, \\
n & \text{:Number of data points}. 
\end{array}\right.
\end{eqnarray}
%
}.
The \(R^2\)-score achieved in this model is $\approx\,0.985$, demonstrating a high level of accuracy in predicting the output variables based on unseen data. 

\section{Model~(2): Informed $K^S$-CNN model}
\label{sec:Model2}
\begin{figure}[!ht]
\begin{center}
\includegraphics[width=12.0cm]{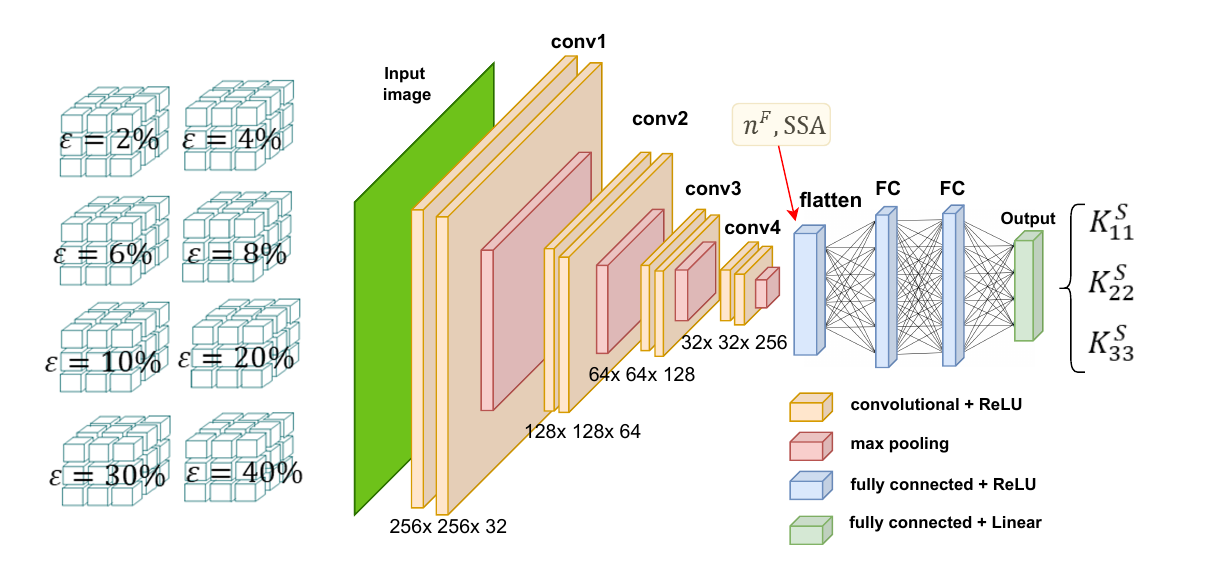}
\end{center}
\caption{Illustration of the ``Informed $K^S$-CNN model'' architecture (see, Fig.~\ref{Fig:CNN_StrucIni} for comparison). This illustration shows the information flow and the inclusion of the physical information ($n^F$ and SSA) through an additional input branch.}
\label{Fig:CNN_StrucIninformed}
\end{figure}

Following the study of \citet{WU2018CNN_Permeability_Synthatic} to improve the accuracy of the CNN model, in this section, we test an alternative structure to our $[K^S\!-\!n^F]$-CNN model presented in Section~\ref{sec:Model1}. In the modified structure, physical parameters, i.e., the porosity ($n^F$) and the  specific surface area (SSA), are incorporated directly into the network architecture instead of having $n^F$ in the loss function.
This results in a kind of physical-parameter-informed CNN, which we  call ``Informed $K^S$-CNN model''.
The modified structure of the CNN model is illustrated in Fig.~\ref{Fig:CNN_StrucIninformed}.

Specifically, an additional input branch (parallel branch) is added to the network to include $n^F$ and SSA, which are passed through a dense layer with 16 units and \textit{ReLU} activation. 
Thus, the output of the Flatten layer (main branch) and the added parallel branch are concatenated to combine the information from both sources. Specifically, the concatenation layer merges the 1D feature vector from the image data and the supplementary features ($n^F$ and SSA), allowing the model to exploit both image-based and additional data.  
As $n^F$ is not anymore a part of the output data, a modified MSE loss function to be considered, which is expressed as
\begin{equation}\label{eq:infLossMSE}
{\cal D}_\text{MSE}=\frac{1}{n}\sum_{i=1}^n\Big[\big(K^{S,p}_{kk,i}-K^{S,t}_{kk,i}\big)^2\Big]\,,    
\end{equation}
where $n$ is the number of output data points and $k=1,2,3$\,.

\begin{figure}[!ht]
\begin{center}
\includegraphics[width=7.5cm]{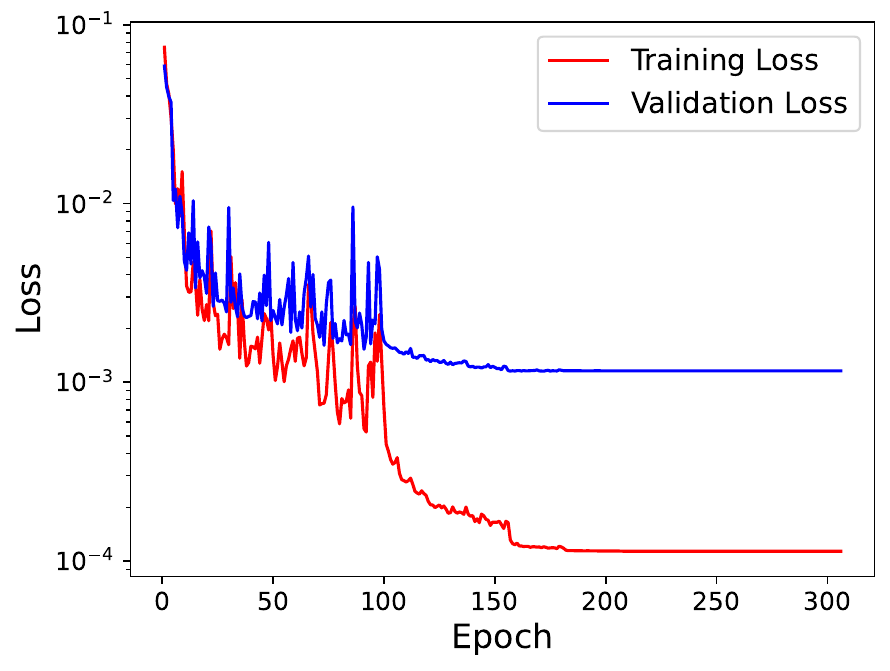}
\end{center}
\caption{Model~(2): Loss function value over the number of weight updates (Epoch).}
\label{Fig:lossEpocModel2}
\end{figure}

The training results of the ``Informed $K^S$-CNN model'' are evaluated through the loss vs. epoch curves, illustrated in Fig.~\ref{Fig:lossEpocModel2}.
These show an initial rapid learning trend. However, similar to $[K^S\!-\!n^F]$-CNN model, the validation loss begins to diverge from the training loss, indicating overfitting. Therefore, the model's state after approximately 100 epochs, corresponding to a loss value of around \( 10^{-3} \), was considered optimal for further evaluation.

\begin{figure}[!ht]
\begin{center}
\includegraphics[width=7.0cm]{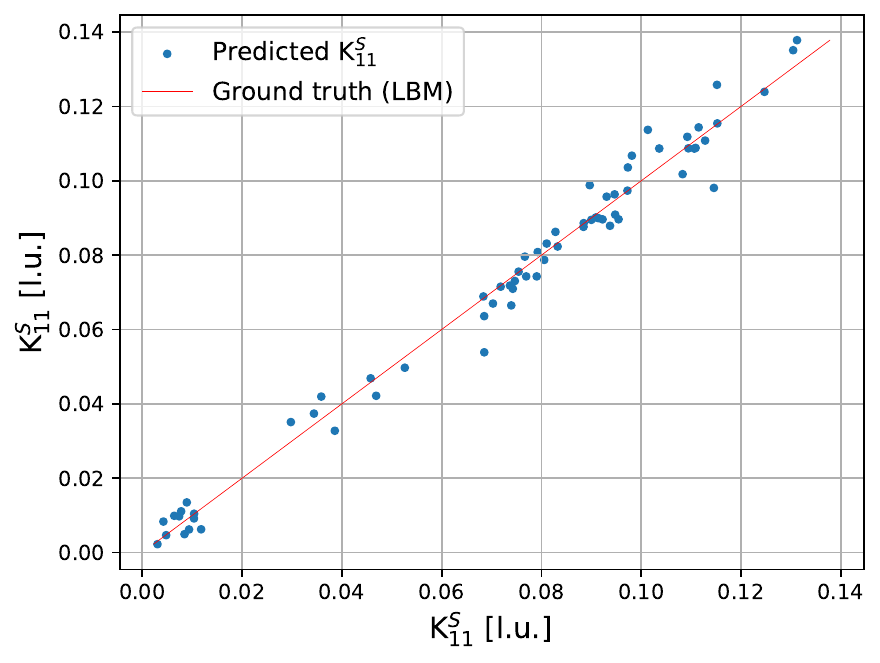}
\includegraphics[width=7.0cm]{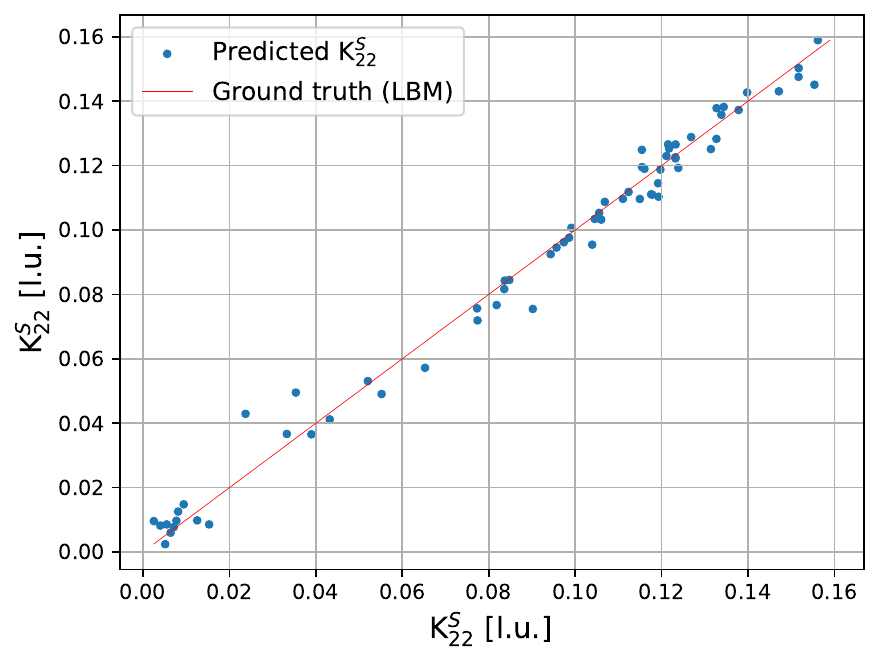}
\includegraphics[width=7.0cm]{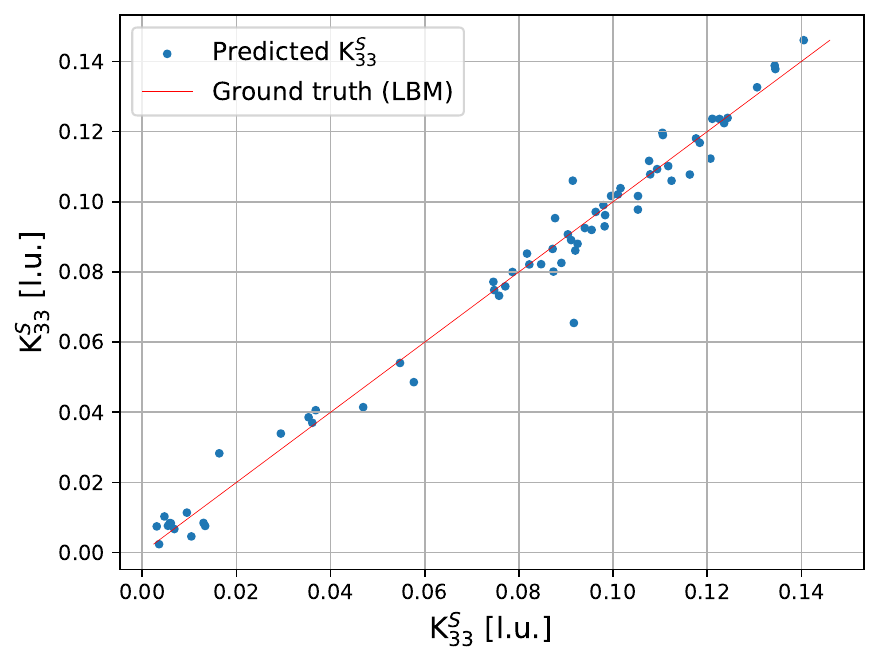}
\end{center}
\caption{Model~(2): CNN-predictions vs. ground-truth values of the intrinsic permeability components.}
\label{Fig:Model2TruePred}
\end{figure}

The scatter plots in Fig.~\ref{Fig:Model2TruePred}, which compare the predicted values to the LBM-based ground truth values for the output variables \(K^{S}_{kk}\) (for \(k=1,2,3\)), further demonstrate the ability of the model to accurately predict the intrinsic permeability values.

The performance of the  ``Informed $K^S$-CNN model'' is quantitatively assessed using the \( R^2 \)-score. The model achieved an \( R^2 \)-score of 0.983, which indicates a high level of accuracy. 
Although the structure of Model~(2) is more complicated than that of Model~(1), its \( R^2 \) score is very close to that of Model~(1). Thus, there is no significant advantage to using Model~(2) over Model~(1) in this particular model with the considered dataset.
Based on this, further research on other datasets is needed to explore the potential benefits of integrating additional physical parameters into neural network architectures.


 \section{Model~(3): Enriched $[K^S\!-\!n^F]$-CNN model}
\label{sec:Model3}
As an alternative to the approach in Section~\ref{sec:Model2}, we explore enhancing the CNN model's performance and generalization through data augmentation, which increases the size and diversity of the training dataset. This improvement follows two key steps:
(1) Training an initial model on synthetic data to capture general patterns.
(2) Applying transfer learning, where the pre-trained model is fine-tuned on the original dataset to refine its performance.

In this work, we utilize open-access synthetic data, published by \citet{NguyenEtAl2022_GAN-RL_Sun}, to enhance our CNN model's predictions of permeability and porosity. The data were generated using Generative Adversarial Networks (GANs), coupled with a Pore Network Model (PNM) to produce visually and physically realistic 3D microstructures of Bentheim sandstone. 
Specifically, GANs are ML algorithms that were originally created in the realm of computer vision to produce synthetic images. GANs comprise a pair of neural networks (NNs), i.e. a generator and a discriminator. These NNs are trained concurrently in a game-theory-inspired framework to generate synthetic datasets that resemble the source datasets, see, e.g. \citet{GoodfellowEtAl2020}.
On the other hand, Pore Network Models (PNMs) are simplified representations of porous media that focus on the structure and connectivity of the pores and throats within the material \cite{Gostick_OpenPNM_a_pore_2016}. PNMs are used to simulate fluid flow and transport properties in porous materials in a simplified way compared to the LBM.
In \cite{NguyenEtAl2022_GAN-RL_Sun}, GANs are coupled with the PNMs using reinforcement learning (RL) to efficiently generate synthetic but realistic 3D images of porous media by incorporating physical constraints and properties derived from the PNMs. 

\begin{figure}[!ht]
\begin{center}
\includegraphics[width=13.0cm]{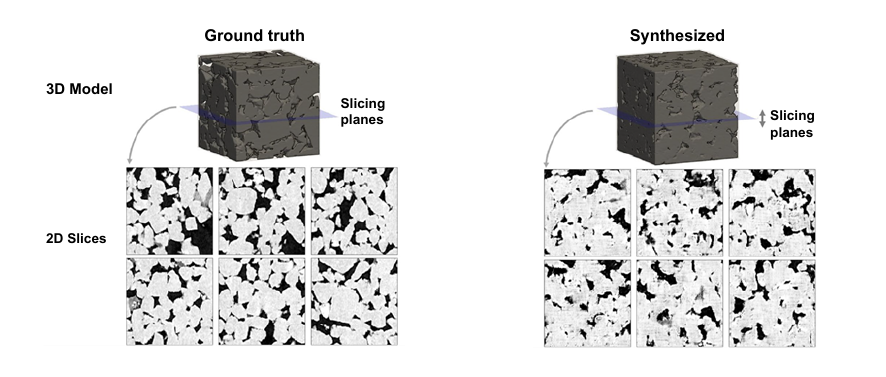}
\end{center}
\caption{2D slices and 3D volume representations of real microstructures with those generated by GANs. The synthetic microstructures exhibit not only high visual fidelity, but also controlled properties such as porosity, permeability, and specific surface area (Source: \citet{NguyenEtAl2022_GAN-RL_Sun}). }
\label{Fig:SynthesizedData}
\end{figure}

Focusing on Bentheim sandstone,  Fig.~\ref{Fig:SynthesizedData} illustrates the comparison between the ground truth and the generated microstructures, which exhibit controlled properties such as porosity, permeability, and specific surface area \cite{NguyenEtAl2022_GAN-RL_Sun}. 

\subsection{Training with the synthetic data}
\label{sec:Model3-TrainingSynData}

The open-access dataset we use for later processing comprises 10 3D volumes, each with dimensions of \(216 \times 216 \times 128\) voxels. During the dataset preparation for training, the gray-scaled synthetic microstructures are first binarized. The binarization threshold is selected to ensure that the resulting porosity matches that of the ground truth. Subsequently, the dataset is sampled, producing a final database consisting of 80 3D samples, each measuring \(108 \times 108 \times 108\) voxels.
Following this, LBM simulations as explained in Section~\ref{OneFluid_LBMTheor} are applied to each of the 3D volumes. To simplify this case of study, uni-directional flow is considered, so that the permeability in the flow direction is computed, i.e. $K^S_{11}$.
Thus, the loss function of this model is expressed as
\begin{equation}
\label{eq:EnrichedLossMSE}
{\cal D}_\text{MSE}=\frac{1}{n}\sum_{i=1}^n\Big[\big(n^{F,p}_{i}-n^{F,t}_{i}\big)^2+\big(K^{S,p}_{11,i}-K^{S,t}_{11,i}\big)^2\Big]\,,   
\end{equation}
where $n$ is the number of output data points.

\begin{figure}[hbt]
\begin{center}
\includegraphics[width=7.5cm]{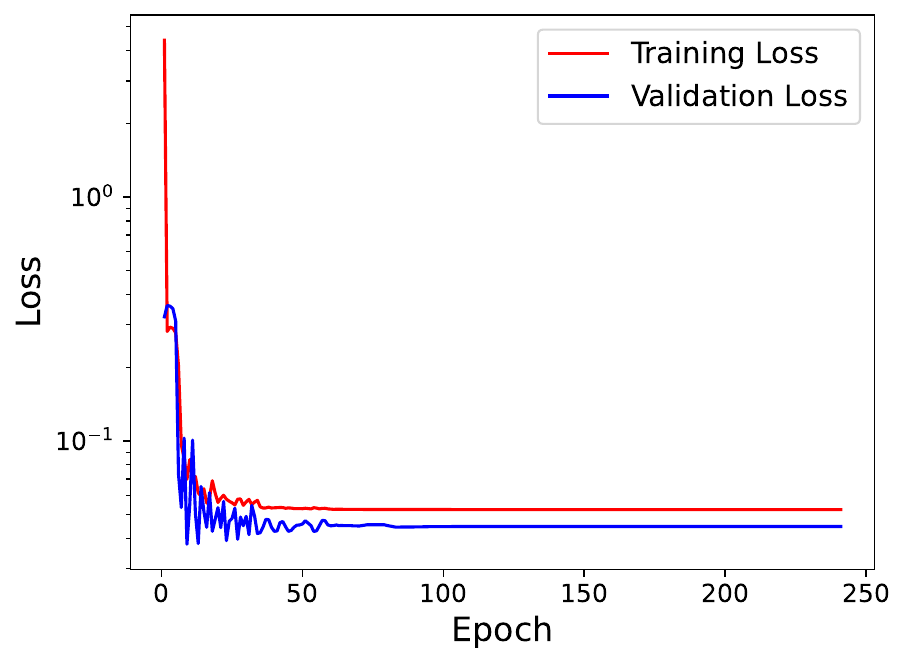}
\end{center}
\caption{Model~(3): Loss function value over the number of weight updates (Epoch) with training on synthetic data.}
\label{Fig:lossEpocModel3}
\end{figure}

%
The training process is carried out using the same model architecture and hyperparameters discussed in Section~\ref{sec:Mode1Arch}.
The training is monitored by evaluating both the training and validation losses over epochs, with the corresponding curves shown in Fig.~\ref{Fig:lossEpocModel3}. 
Although the database is relatively small, both losses decrease rapidly, indicating that the model is learning effectively. As training continues, the training loss stabilizes at a slightly higher value compared to the validation loss, suggesting that the model is encountering some degree of underfitting. Both values of the loss function are however close to $5\times10^{-2}$.
The weights and biases of the trained model are stored in $\ast$.hdf5 file, which is then used for training on the original data within the context of transfer learning.

\subsection{Transfer learning effect on the model learning}
Transfer learning in our CNN model leverages the pre-trained model's knowledge (weights and biases) from Section~\ref{sec:Model3-TrainingSynData} to enhance learning efficiency and performance on the real dataset. This dataset consists of 448 3D volumes, each with dimensions of \(108 \times 108 \times 108\) voxels. Uni-directional flow simulations using the LBM are then conducted to generate the database, which includes permeability and porosity as outputs, as detailed in Section~\ref{sec:Database}.
The CNN model architecture and hyperparameters applied are identical to those described in Section~\ref{sec:Mode1Arch}, and the loss function is defined by Eq.~(\ref{eq:EnrichedLossMSE}). 
\begin{figure}[!ht]
\begin{center}
\includegraphics[width=7.0cm]{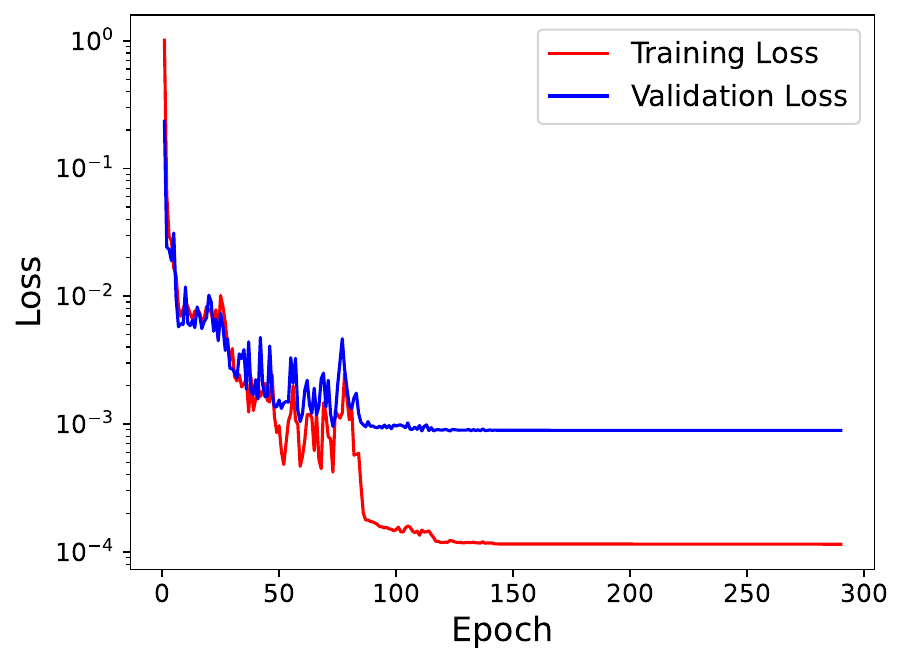}
\includegraphics[width=7.0cm]{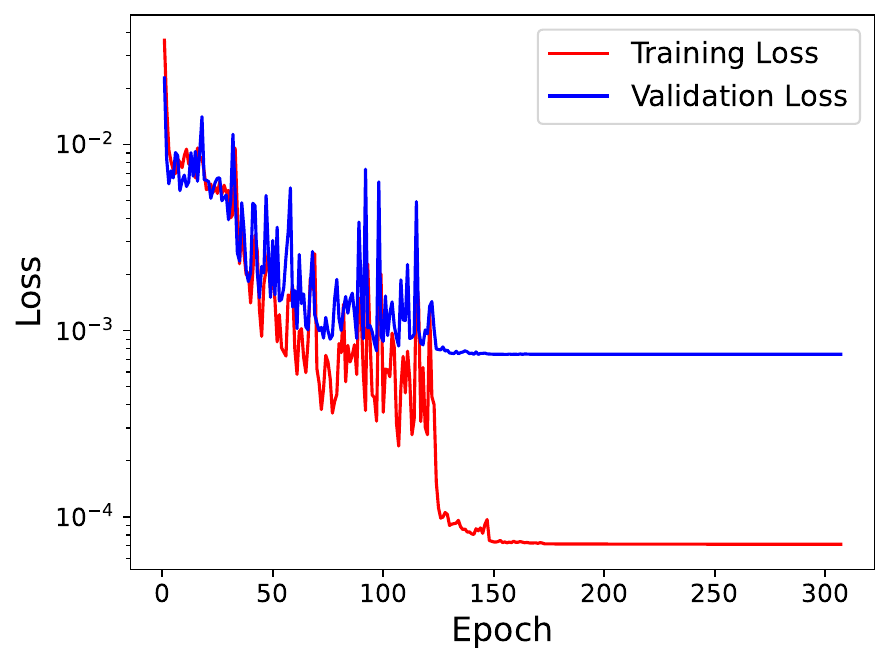}
\end{center}
\caption{Model~(3): Loss function value over the number of weight updates (Epoch) with no transfer learning (left) and with transfer learning (right).}
\label{Fig:lossEpocModel3TransferLearning}
\end{figure}

In this study, we compare two models: one without transfer learning (random initial weights) and one with transfer learning (initial weights from the pre-trained model on synthetic data).
The results, illustrated in Fig.~\ref{Fig:lossEpocModel3TransferLearning}, demonstrate that the model with transfer learning exhibits faster learning compared to the model without transfer learning. Notably, the initial loss value is around 1 for the model without transfer learning, while it is approximately $0.03$ for the model with transfer learning. After sufficient epochs (around 300), the final values of the loss functions for both models are similar.


In summary, transfer learning significantly enhances the learning efficiency of the CNN model when applied to the original dataset. The pre-trained weights and biases facilitate quicker convergence, thereby reducing the computational resources and time required to achieve optimal performance.
Thus, we conclude that the pre-trained weights and biases facilitate faster convergence and help reduce the computational resources and time required to achieve a well-trained model.
To demonstrate the importance and effectiveness of this approach, further testing is needed, especially with smaller real datasets. In such cases, the generation of synthetic data and transfer learning could be crucial for the development of a well-trained ML model with high predictive power.

\section{Conclusions and future aspects}
\label{sec:Conclusions}

In this work, we have demonstrated the ability of different CNN models to accurately predict the anisotropic intrinsic permeability tensor together with the porosity at different deformation states. The inputs of this CNN model are real CT images related to Bentheim sandstones. The predicted parameters (outputs) can be directly integrated into a macroscopic TPM framework to describe both linear and non-linear flow through porous media, where different flow models have been discussed.
Three CNN models are presented in this study: (1) An initial model that includes the permeability components and porosity in the definition of the loss function.
(2) An informed CNN model that incorporates physical parameters directly into the model architecture.
(3) An enriched CNN model with transfer learning, where learning starts with synthetic data and the pre-trained model is fine-tuned on the original dataset.
The initial CNN model showed high accuracy, while the informed CNN model did not significantly outperform the simpler initial model on the given dataset. However, data augmentation with synthetic data and the application of transfer learning proved effective in improving learning efficiency.

Future research will focus on several promising areas to further improve and extend current models. 
This includes exploring the potential of using CNN models for the inverse design of porous metamaterials with tailored hydro-mechanical properties, such as stiffness and permeability, that can be fabricated by additive manufacturing.
Another future approach is Physics-Enhanced Multiscale Modeling, where physics-based constraints and principles are integrated into the multiscale modeling framework. This approach could potentially allow the use of smaller data sets, improving the efficiency and feasibility of training accurate models without compromising performance.

 \section*{Conflict of interest}
 The authors declare that there is no conflict of interest.


\begin{acknowledgements}
Y. Heider would like to gratefully thank the  German Research Foundation (DFG) for the support of the project  ``Multi-field continuum modeling of two-fluid-filled porous media fracture augmented by microscale-based machine-learning material laws'', grant number 458375627. F. Aldakheel gratefully acknowledges support for this research by the “German Research Foundation” (DFG) through the SFB/TRR-298-SIIRI—Project-ID 426335750.
\end{acknowledgements}

\section*{Data availability statement}
Exemplary source codes are made available as open-access for interested readers at [\href{https://doi.org/10.25835/xrii0m6f}{https://doi.org/10.25835/xrii0m6f}].
%
%

\bibliographystyle{spbasic}      
\bibliography{main}
%
%

\end{document}